\documentclass[12pt]{article}
\usepackage{amsmath,amssymb}
\usepackage{bm}
\usepackage{graphicx,color}
\usepackage{wrapfig}

\begin{document}

\begin{flushright}
KEK-TH-1830
\end{flushright}

\vspace{2cm}

\begin{center}

{\bf \LARGE Sensitivity of  High-Scale SUSY \\ \vspace{0.3cm}in Low Energy Hadronic FCNC} 
\\

\vspace*{1.5cm}
{\large 
Morimitsu~Tanimoto$^{a,}$\footnote{E-mail address: tanimoto@muse.sc.niigata-u.ac.jp} \ \ and \ \
Kei~Yamamoto$^{b,}$\footnote{E-mail address: kei.yamamoto@kek.jp}
} \\
\vspace*{1.5cm}

{\it
$^b$Department of Physics, Niigata University,\\
Niigata 950-2181, Japan\\
$^b$Institute of Particle and Nuclear Studies,\\
High Energy Accelerator Research Organization (KEK),\\
Tsukuba 305-0801, Japan\\
}

\end{center}

\vspace*{0.5cm}

\begin{abstract}
{\normalsize 
 We discuss the sensitivity of the high-scale SUSY at $10$-$1000$ TeV in 
  $B^0$, $B_s$, $K^0$ and $D$  meson systems together with the neutron EDM and the mercury EDM.
  In order to estimate the contribution of the squark flavor mixing to these FCNCs,
we  calculate the squark mass spectrum, which is consistent with
   the recent Higgs discovery. 
 The SUSY contribution in $\epsilon_K$ could be large, around $40\%$ in the region
of  the SUSY scale   $10$-$100$ TeV.
The neutron EDM and the mercury EDM are also sensitive to the SUSY contribution induced by
the gluino-squark interaction. 
The predicted  EDMs are roughly proportional to $|\epsilon_K^{\rm SUSY}|$.
If the SUSY contribution is the level of ${\cal O}(10\%)$ for $\epsilon_K$,
the neutron EDM is expected to be discovered in the region of $10^{-28}$-$10^{-26}$ecm.
The mercury EDM also gives a strong constraint for the gluino-squark interaction. 
The SUSY contribution of $\Delta M_D$ is also discussed.
}
\end{abstract} 

\newpage
%%%%%%%%%%%%%%%%%%%%%%%%%%%%%%%%%%%%%%%%%%%%%%%%%%%%%%%%%%%
%%%%%%%%%%%%%%%%%%%%%%%%%%%%%%%%%%%%%%%%%%%%%%%%%%%%%%%%%%%
\section{Introduction}
%%%%%%%%%%%%%%%%%%%%%%%%%%%%%%%%%%%%%%%%%%%%%%%%%%%%%%%%%%%
%%%%%%%%%%%%%%%%%%%%%%%%%%%%%%%%%%%%%%%%%%%%%%%%%%%%%%%%%%%
\label{sec:Intro}

The supersymmetry (SUSY) is one of the most attractive theories beyond the standard model (SM).
Therefore, the SUSY has been expected to be observed at the LHC experiments.
However, no signals of the SUSY  have been discovered yet.
 The  present searches for the SUSY particles give us important constraints for the SUSY.
Since the lower bounds of the superparticle masses increase gradually, 
the squark and the gluino masses are supposed  
to be at the higher scale than $1$ TeV ~\cite{Aad:2014wea,Chatrchyan:2014lfa,Aad:2014kra}. 
On the other hand, the SUSY model has been seriously constrained  by the Higgs discovery, in which
the Higgs mass is  $125$ GeV~\cite{Higgs}. 
Based on  this theoretical and experimental situations, 
  we consider the high-scale SUSY models, which  have been widely 
discussed with a lot of attention
\cite{Giudice:1998xp}-\cite{Nagata:2015hha}.

If the squark and slepton masses
are at the high-scale  ${\cal O}(10$-$1000)$ TeV,
 the lightest Higgs mass can be pushed up to $125$ GeV, whereas  SUSY particles are out of the reach of the LHC experiment.
Therefore, the indirect search of the SUSY particles becomes important in the low energy
flavor physics \cite{Altmannshofer:2013lfa,Moroi:2013sfa,McKeen:2013dma}.

The flavor physics is also  on the new stage in the light of LHCb data.
The LHCb collaboration has reported 
new data of the CP violation of the $B_s$ meson and the branching ratios 
of rare $B_s$ decays~\cite{Bediaga:2012py}-\cite{LHCb:2011ab}.
For many years the CP violation in the $K$ and $B^0$ mesons 
has been successfully understood within the framework of the standard model (SM), 
so called Kobayashi-Maskawa (KM) model \cite{Kobayashi:1973fv}, 
where the source of the CP violation is the KM phase 
in the quark sector with three families. 
However, the new physics has been expected to be indirectly discovered
in the precise data of $B^0$ and $B_s$ meson decays at the LHCb experiment and the further coming  experiment, Belle-II. 

There are new sources of the CP violation if the SM is 
extended to the SUSY models. The soft squark mass matrices contain 
the CP violating phases, which contribute to the flavor changing 
neutral current (FCNC) with the CP violation \cite{Gabbiani:1996hi}. 
Therefore, we can expect the SUSY effect in the CP violating phenomena. 
However, the clear deviation from the  SM prediction
has not been observed yet in the LHCb experiment~\cite{Bediaga:2012py}-\cite{LHCb:2011ab}.
Actually, we have found that the CP violation of $B^0$ and $B_s$ meson systems
are suppressed if the SUSY scale is above $10$ TeV \cite{Tanimoto:2014eva}.
On the other hand, the CKMfitter group presented the current limits on new physics contributions  of ${\cal O}(10\%)$  in  $B^0$, $B_s$  and $K^0$ systems  \cite{Charles:2013aka}.
They have also estimated the sensitivity to new physics in  $B^0$ and $B_s$ mixing 
achievable with  $50\rm ab^{-1}$ of Belle-II and $50\rm fb^{-1}$ of LHCb data.
Therefore, we should carefully study the sensitivity of the high-scale SUSY to the hadronic FCNC.

In this work, we discuss  the high-scale SUSY contribution to the   $B^0$, $B_s$ and $K^0$ meson systems. Furthermore, we  also discuss the sensitivity to the $D$ meson
and the electric dipole moment (EDM) of the neutron and the  mercury.
For these modes,  the most important process of the SUSY contribution  is
 the gluino-squark mediated  flavor changing process \cite{King:2010np}- \cite{Hayakawa:2013dxa}.
 The CP violation of $K$ meson, $\epsilon_K$,   provides a severe constraint to the gluino-squark mediated FCNC \cite{Mescia:2012fg,Tanimoto:2015ota}. 
In addition, 
the recent work have found  that the chromo-electric dipole moment (cEDM) is sensitive to the high-scale SUSY \cite{Fuyuto:2013gla}.
It is noted   that the upper-bound of the neutron EDM (nEDM) \cite{PDG}
gives a  severe constraint for
  the gluino-squark interaction through the cEDM \cite{Pospelov:2000bw}-\cite{Fuyuto:2012yf}.
It is also remarked that the upper bound of the mercury EDM (HgEDM)  \cite{Griffith:2009zz}
can give an important   constraint \cite{Chiou:2006qk}.

%%%%%%%%%%%%%%%%%%%%%%%%%%%%%%%%%%%%%%%
 In order to estimate the gluino-squark mediated FCNC 
  of the  $K$, $B^0$, $B_s$ and $D$ mesons,
  we work in the basis of  the squark mass eigenstate
 with the non-minimal squark (slepton) flavor mixing. 
 There are three reasons why the SUSY contribution to the FCNC considerably depends on the squark mass spectrum.
  The first one is that the GIM mechanism works in the squark flavor mixing, and 
  the second one is that the loop functions depend on the mass ratio of squark and gluino.
  The last one is that we need the mixing angle between the left-handed sbottom and right-handed sbottom,
  which dominates the $\Delta B=1$ decay processes.
   Therefore, we discuss the squark mass spectrum, which is consistent with
   the recent Higgs discovery.
   Taking  the universal soft parameters at the SUSY breaking scale, we obtain
   the squark mass spectrum at the matching scale where
 the SM emerges, by using the Renormalization Group Equations (RGEs) of the soft masses.
  On the other hand, the $6\times 6$ mixing matrix between  squarks and quarks is 
  taken to be free at the low energy.

In section 2, we discuss the squark and gluino mass spectrum and the squark mixing.
In section 3, we present the formulation of the  FCNC with $\Delta F=2$ in 
$K$, $B^0$, $B_s$ and $D$ meson systems together with nEDM and HgEDM. 
We present  numerical results and discussions in section 4.
Section 5 is devoted to the summary. 
The  relevant formulations are  presented in Appendices A, B, C and D.

%%%%%%%%%%%%%%%%%%%%%%%%%%%%%%%%%%%%%%%%%%%%%%%%%%%%%%%%%%%
%%%%%%%%%%%%%%%%%%%%%%%%%%%%%%%%%%%%%%%%%%%%%%%%%%%%%%%%%%%
\section{SUSY Spectrum and Squark mixing}
%%%%%%%%%%%%%%%%%%%%%%%%%%%%%%%%%%%%%%%%%%%%%%%%%%%%%%%%%%%
%%%%%%%%%%%%%%%%%%%%%%%%%%%%%%%%%%%%%%%%%%%%%%%%%%%%%%%%%%%
\label{sec:Pre}
The low energy FCNCs depend significantly on the spectrum of the SUSY particles,
which depend on the model.
As well known, the lightest Higgs mass can be pushed up to $125$ GeV
if the squark masses are expected to be ${\cal O}(10)$ TeV. 
Therefore,
let us consider  the heavy SUSY particle mass spectrum in the framework of
the minimal supersymmetric standard model (MSSM), which is consistent with the observed Higgs mass.
The discussion how to obtain the SUSY spectrum have been given in Refs.
\cite{Delgado:2013gza, Giudice:2006sn}.

We outline how to obtain the SUSY spectrum in our work. The details are presented in Appendix A.
At the SUSY breaking scale $\Lambda$, we write the quadratic terms in the MSSM potential  as
\begin{equation}
 V_2=m_1^2 |H_1|^2+ m_2^2 |H_2|^2+m_3^2 (H_1\cdot H_2+ h.c.) \ .
\end{equation}
Then, the Higgs mass parameter $m^2$ is expressed in terms of $m_1^2$, $m_2^2$ and $\tan\beta$
as:
\begin{eqnarray}
m^2=\frac{m_1^2-m_2^2\tan^2\beta}{\tan^2\beta -1} \ .
\end{eqnarray}
After running down to  the  $Q_0$ scale, in which the SM emerges, 
by the one-loop SUSY Renormalization Group Equations (RGEs) \cite{Martin:1997ns},
the scalar potential is  the SM one as follows:
\begin{equation}
 V_{SM}=-m^2 |H|^2+\frac{\lambda}{2} |H|^4 \ .
\end{equation}
Here, the Higgs coupling $\lambda$ is given in terms of the SUSY parameters
at the leading order as
 \begin{equation}
 \lambda(Q_0)=\frac{1}{4} (g^2+g'^2)\cos^2 2\beta +\frac{3h_t^2}{8\pi^2} X_t^2 \left (1- \frac{X_t^2}{12}\right ) \ , \qquad
 X_t=\frac{A_t(Q_0)-\mu(Q_0)\cot\beta}{ Q_0 } \ ,
\end{equation}
and $h_t$ is the top Yukawa coupling of the SM.
 The parameters $m_2$ and $\lambda$ run with the two-loop SM RGEs with ${\rm \overline{MS}}$ scheme
 \cite{Iso:2012jn} down to the electroweak scale $Q_{EW}=m_H$, and then give
\begin{equation}
 m_H^2=2m^2(m_H)=\lambda(m_H)v^2\ .
\end{equation}

When $m_H=125$ GeV is put,  $\lambda(Q_0)$ and $m^2(Q_0)$ are obtained. This  input constrains the SUSY mass spectrum of the MSSM.
In our work, we take the  universal soft breaking parameters at the SUSY breaking scale $\Lambda$ as follows:
\begin{eqnarray}
&&m_{\tilde Q_i}(\Lambda)=m_{\tilde U^c_i}(\Lambda)=m_{\tilde D^c_i}(\Lambda)=
m_{\tilde L_i}(\Lambda)=m_{\tilde E^c_i}(\Lambda)=m_0^2 \ (i=1,2,3) \ , \nonumber \\
&&M_1(\Lambda)=M_2(\Lambda)=M_3(\Lambda)=m_{1/2}  \ , \qquad 
m_1^2({\Lambda})= m_2^2({\Lambda})=m_0^2 \ , \nonumber \\
&&A_U({\Lambda})=A_0 y_U(\Lambda)\ , \quad A_D({\Lambda})=A_0 y_D(\Lambda)\ , 
\quad A_E({\Lambda})=A_0 y_E(\Lambda)\ .
\end{eqnarray}
By inputting   $m_H=125$ GeV and taking the heavy scalar mass $m_{{\cal H}}\simeq Q_0$
(see Appendix A), 
we can obtain the SUSY spectrum for the fixed $Q_0$ and $\tan\beta$.
The details and numerical results are presented in Appendix A.

%%%%%%%%%%%%%%%%%%%%%%%%%%%%%%%%%%%%%%%%%%%%%%%%%
%%%%%%%%%%%%%%%%%%%%%%%%%%%%%%%%%%%%%%%%%%%%%%%%%
Let us consider the squark flavor mixing.
As discussed above, there is no flavor mixing at $\Lambda$ in the MSSM.  
However, in order to consider the non-minimal flavor mixing framework, we allow the off diagonal components of the squark mass matrices at the $10\%$ level, which leads to the
flavor mixing of order $0.1$.
 We take these flavor mixing angles as free parameters  at low energies.
Now we consider the $6\times 6$ squark mass matrix  $M_{\tilde q}$ in the super-CKM basis.
In order to move the mass eigenstate basis of squark masses,
we should diagonalize the mass matrix by rotation matrix $\Gamma _{G}^{(q)}$  as  
\begin{equation}
m_{\tilde q}^2=\Gamma _{G}^{(q)}M_{\tilde q}^2 \Gamma _{G}^{(q)  \dagger} \ ,
\end{equation}
where $\Gamma _{G}^{(q)}$ is the $6\times 6$ unitary matrix, and we decompose it into the $3\times 6$ matrices 
 as $\Gamma _{G}^{(q)}=(\Gamma _{GL}^{(q)}, \ \Gamma _{GR}^{(q)})^T$ in the following expressions:
\begin{align}
\Gamma _{GL}^{(d)}&=
\begin{pmatrix}
c_{13}^L & 0 & s_{13}^L e^{-i\phi_{13}^L} c_{\theta} & 0 & 0 & -s_{13}^L e^{-i\phi_{13}^L} s_{\theta} e^{i \phi} \\
-s_{23}^L s_{13}^L e^{i(\phi_{13}^L-\phi_{23}^L)} & c_{23}^L & s_{23}^Lc_{13}^Le^{-i\phi_{23}^L}c_{\theta} & 0 & 0 & -s_{23}^Lc_{13}^Le^{-i\phi_{23}^L}s_{\theta}e^{i\phi} \\
-s_{13}^Lc_{23}^Le^{i\phi_{13}^L} & -s_{23}^Le^{i\phi_{23}^L} &c_{13}^Lc_{23}^Lc_{\theta} & 0 & 0 & 
-c_{13}^Lc_{23}^Ls_\theta e^{i\phi}
\end{pmatrix}, \nonumber \\
\nonumber\\
\Gamma _{GR}^{(d)}&=
\begin{pmatrix}
0 & 0 & s_{13}^R s_\theta e^{-i \phi_{13}^R} e^{-i\phi } & c_{13}^R & 0 & s_{13}^R e^{-i \phi_{13}^R} c_{\theta} \\
0 & 0 & s_{23}^R c_{13}^R s_\theta e^{-i \phi_{23}^R} e^{-i\phi } & -s_{13}^R s_{23}^R e^{i(\phi_{13}^R-\phi_{23}^R)} & c_{23}^R & s_{23}^R c_{13}^R e^{-i \phi_{23}^R} c_{\theta} \\
0 & 0 & c_{13}^R c_{23}^R s_{\theta} e^{-i\phi } & -s_{13}^R c_{23}^R e^{i\phi_{13}^R} & -s_{23}^R e^{i \phi_{23}^R} & 
c_{13}^R c_{23}^R c_{\theta} \\
\end{pmatrix},
\label{mixing}
\end{align} 
where we use abbreviations $c_{ij}^{L,R}=\cos\theta_{ij}^{L,R}$, $s_{ij}^{L,R}=\sin\theta_{ij}^{L,R}$, 
 $c_\theta =\cos \theta$ and $s_\theta =\sin \theta$.
Here $\theta$ is  the left-right mixing angle  between   $\tilde b_{L}$ and $\tilde b_{R}$,
which is discussed in Appendix A.
 It is remarked that  we take $s_{12}^{L,R}=0$ due to the degenerate squark masses of the first and second families as discussed in  Appendix A. 

The gluino-squark-quark interaction is given as
\begin{equation}
\mathcal{L}_\text{int}(\tilde gq\tilde q)=-i\sqrt{2}g_s\sum _{\{ q\} }\widetilde q_i^*(T^a)
\overline{\widetilde{G}^a}\left [(\Gamma _{GL}^{(q)})_{ij}{L}
+(\Gamma _{GR}^{(q)})_{ij}{R}\right ]q_j+\text{h.c.}~,
\end{equation}
where $L=(1-\gamma _5)/2$, $R=(1+\gamma _5)/2$, and $\widetilde G^a$ denotes the gluino field,  
$q^i$ are three left-handed (i=1,2,3) and three right-handed quarks (i=4,5,6).
This interaction leads to the gluino-squark mediated flavor changing processes 
with $\Delta F=2$ and  $\Delta F=1$ through the  box and  penguin diagrams.

The chargino (neutralino)-squark-quark  interaction can be also
 discussed in the similar way.

%%%%%%%%%%%%%%%%%%%%%%%%%%%%%%%%%%%%%%%%%%%%%%%%%%%%%%%%%%%
%%%%%%%%%%%%%%%%%%%%%%%%%%%%%%%%%%%%%%%%%%%%%%%%%%%%%%%%%%%
\section{FCNC of $\Delta F=2$}
%%%%%%%%%%%%%%%%%%%%%%%%%%%%%%%%%%%%%%%%%%%%%%%%%%%%%%%%%%%
%%%%%%%%%%%%%%%%%%%%%%%%%%%%%%%%%%%%%%%%%%%%%%%%%%%%%%%%%%%
\label{sec:Setup}
In our previous work \cite{Tanimoto:2014eva}, 
  we  have probed the high-scale SUSY, which is at $10$-$50$ TeV scale, 
in the CP violations of $K$, $B^0$ and $B_s$ mesons.
It is found that $\epsilon_K$ is most sensitive to the SUSY even if the SUSY scale is at $50$ TeV.
The SUSY contributions for the  time dependent CP asymmetries of $B^0$ and $B_s$ 
with $\Delta B=1$ are suppresses at the SUSY scale of $10$ TeV.
Furthermore, the SUSY contribution for the $b\to s\gamma$ process is also 
suppressed since the left-right mixing angle, which induces the chiral enhancement,
  is very small as discussed in Appendix A.
Therefore, we
 discuss the neutral meson mixing $P^0$-$\bar P^0 (P^0=K, B^0, B_s, D)$,
which  are  FCNCs with  $\Delta F=2$.

In those FCNCs, the dominant SUSY contribution is given through  the gluino-squark interaction.
Then,  the dispersive part of meson mixing $M_{12}^{P^0}(P^0=K, B^0, B_s)$ are written  as 
\begin{equation}
M_{12}^{P^0}=M_{12}^{P^0,\text{SM}}+M_{12}^{P^0,\text{SUSY}},
\end{equation}
where $M_{12}^{q,\text{SUSY}}$ are given by the squark mixing parameters in Eq.(\ref{mixing}) and its explicit formulation is given in Appendices B and C.

At first, we discuss the  $\Delta B=2$ process, that is, 
the mass differences $\Delta M_{B^0}$ and $\Delta M_{B_s}$, 
and the CP-violating phases $\phi_d$ and $\phi_s$.
In general, the contribution of the new physics (NP) to the dispersive part $M_{12}^q$ 
is parameterized as 
\begin{equation}
M_{12}^{B_q}=M_{12}^{q,\text{SM}}+M_{12}^{q,\text{NP}}=
M_{12}^{q,\text{SM}}(1+h_qe^{2i\sigma _q})~, \quad (q=B^0,B_s)
\label{NP}
\end{equation}
where $M_{12}^{q,\text{NP}}$ are the NP contributions.
The generic fits for $B^0$ and $B_s$ mixing have given the constraints on $(h_q, \sigma_q)$
\cite{Charles:2013aka}, where
it is assumed that the NP does not significantly affect 
the SM tree-level charged-current interaction, that is,  the absorptive part $\Gamma_{12}^q$ is dominated by the decay $b\to c\bar c s$. 
 At present, the NP contribution $h_q$  are $10$-$35\%$ and $15$-$25\%$ depending on $\sigma_q$
 for  $B^0$ and $B_s$, respectively.
 Thus, we can expect the sizable NP contribution of ${\cal O}(20\%)$.
 We will discuss whether the high-scale SUSY can fill in the  magnitude of the present  NP contribution of ${\cal O}(20\%)$.

%%%%%%%%%%%%%%%%%%%%%%%%%%

Next, we discuss $\Delta S=2$ process,  $\Delta M_{K^0}$ and 
 the CP-violating parameter in the $K$ meson, $\epsilon_K$.
By the similar parametrization in Eq.(\ref{NP}), 
the allowed region of $(h_K, \sigma_K)$ has been estimated in Ref.\cite{Charles:2013aka}.
The NP contribution is at least  $50\%$ although there is the strong $\sigma_K$ dependence. Therefore, it is important to examine carefully  the CP violating
parameter $\epsilon_K$, which is given as follows:
\begin{equation}
\epsilon_K
=
e^{i \phi_{\epsilon}} \sin{\phi_{\epsilon}} \left( \frac{\text{Im}(M_{12}^K)}{\Delta M_K}
+ \xi \right),  \qquad
\xi
=
\frac{\text{Im} A_0^K}{\text{Re} A_0^K} , \qquad
\phi_\epsilon=\tan^{-1}\left( \frac{2 \Delta M_K}{\Delta \Gamma_K} \right),
\end{equation}
with $A_0^K$ being the isospin zero amplitude in $K\to\pi\pi$ decays.
Here, $M_{12}^K$ is the dispersive part of the  $K^0$-$\bar{K^0}$ mixing, 
and $\Delta M_K$ is the mass difference in the neutral $K$ meson.
The effects of $\xi \ne 0$ and $\phi_{\epsilon} < \pi/4$ give suppression effect in $\epsilon_K$,
and it is parameterized as $\kappa_{\epsilon}$ and estimated by Buras and Guadagnoli \cite{Buras:2008nn} as:
\begin{equation}
\kappa_{\epsilon}
=
0.92 \pm 0.02 \ \ .
\end{equation}
In the SM, the dispersive part $M_{12}^K$ is given as follows, 
\begin{align}
M_K^{12}
&=
\langle K| \mathcal{H}_{\Delta F=2} |\bar{K} \rangle \nonumber \\
&=
-\frac{4}{3}\left( \frac{G_F}{4 \pi} \right)^2 M_W^2 \hat{B}_K F_K^2 M_K \left( \eta_{cc} \lambda_c^2 E(x_c)
+\eta_{tt} \lambda_t^2 E(x_t)
+2 \eta_{ct} \lambda_c \lambda_t E(x_c,x_t) \right) ,
\end{align}
where $\lambda_c = V_{cs}V_{cd}^*,\  \lambda_t = V_{ts}V_{td}^*$.
 The $E(x)$'s are the one-loop functions \cite{Inami:1980fz}
and  $\eta_{cc,tt,ct}$ are the QCD corrections \cite{Buras:2008nn}.
Then, $|\epsilon_K^{\text{SM}}|$ is given in terms of the Wolfenstein parameters $\lambda$, $\overline\rho$ and 
$\overline\eta$ as follows:
\begin{align}
|\epsilon_K^{\text{SM}}|
&=
\kappa_{\epsilon} C_{\epsilon} \hat{B}_K |V_{cb}|^2 \lambda^2 \bar{\eta} 
\left( |V_{cb}|^2 (1-\bar{\rho})\eta_{tt} E(x_t)
-\eta_{cc}E(x_c)
+\eta_{ct} E(x_c,x_t) \right)  \ , %\nonumber \\
%&=
%\kappa_{\epsilon} C_{\epsilon} \hat{B}_K |V_{cb}|^2 \lambda^2
%\left(
%\frac{1}{2} |V_{cb}|^2 R_t^2 \sin(2 \beta) \eta_{tt} E(x_t)
%+R_t \sin \beta (-\eta_{cc}E(x_c) +\eta_{ct} E(x_c,x_t))
%\right),
\label{epsilonKSM}
\end{align}
with
\begin{align}
 C_{\epsilon}=
\frac{G_F^2 F_K^2 m_K M_W^2}{6  \sqrt{2} \pi^2 \Delta M_K}.
\end{align}

 Note that $|\epsilon_K^{\text{SM}}|$  depends on the non-perturbative parameter $\hat{B}_K$ in Eq.(\ref{epsilonKSM}). 
Recently, the error of this parameter shrank dramatically in the lattice calculations \cite{Bae:2013lja}.
In our calculation we use the  updated value by the Flavor Lattice Averaging Group \cite{Aoki:2013ldr}:
\begin{equation}
\hat B_K= 0.766 \pm 0.010\ \ .
\label{BK}
\end{equation}

Let us write down   $\epsilon_K$  as:
 \begin{align}
  \epsilon_K=\epsilon_K^{\text{SM}}+\epsilon_K^{\rm SUSY},
  \label{epsilon}
  \end{align}
where $\epsilon_K^{\rm  SUSY}$ is induced by the imaginary part of the gluino-squark box diagram, 
which is presented in Appendices B and C. 
Since $s_{12}^{L(R)}$ vanishes in our scheme, 
 $\epsilon_K^{SUSY}$ is given in the second order of the squark mixing
  $s_{13}^{L(R)}\times s_{23}^{L(R)}$.

%%%%%%%%%%%%%%%%%%%%%%%%%%%%%%%%%%%

In addition to the above FCNC processes, 
the neutron EDM, $d_n$  arises through the cEDM of the quarks, $d_q^C$
 due to  the gluino-squark mixing \cite{Pospelov:2000bw}-\cite{Fuyuto:2012yf}.
 By using the QCD sum rules, $d_n$ is given as 
 \begin{equation}
  d_n =  (0.79 d_d-0.20 d_u) + e(0.3 d_u^C + 0.59 d_d^C) \ .
\label{cedmQCD}
\end{equation}
where  $d_q$ and  $d_q^C$ denote the EDM and cEDM of quarks  $d_q^C$  defined in  Appendix D.
On the other hand, by using the chiral perturbation theory
\begin{equation}
  d_n = e (3.0 d_u^C + 2.5 d_d^C + 0.5 d_s^C) \ .
\label{cedmChiral}
\end{equation}
Therefore,  the  experimental upper bound \cite{PDG}
\begin{equation}
  |d_n| < 0.29 \times 10^{-25}{\rm ecm} \ ,
 \end{equation}
 provides us a  strong constraint to the gluino-squark mixing.

 The HgEDM can also probe the gluino-squark mixing \cite{Chiou:2006qk}.
The QCD sum rule approach gives \cite{Falk:1999tm}
\begin{equation}
  d_{Hg} = e (d_u^C - d_d^C + 0.012 d_s^C)\times 3.2\times 10^{-2} \ ,
\label{cedmHgQCD}
\end{equation}
and the chiral Lagrangian method gives \cite{Hisano:2004pw}
\begin{equation}
  d_{Hg} = e (d_u^C - d_d^C + 0.0051 d_s^C)\times 8.7\times 10^{-3} \ .
\label{cedmHgChiral}
\end{equation}
The experimental upper bound \cite{Griffith:2009zz}
\begin{equation}
  |d_{Hg}| < 3.1 \times 10^{-29}{\rm ecm} \ ,
 \end{equation}
 constrains the gluino-squark mixing.

%%%%%%%%%%%%%%%%%%%%%%%%%%%%%%%%%%%%%%%%%%%%

At the last step, we discuss the  charm sector, which is a promising field to probe for the new physics beyond the SM.  The $D^0-\bar D^0$ mixing is now well established 
\cite{Bevan:2014tha} as follows:
 \begin{align}
  x_D=\frac{\Delta M_D}{\Gamma_D}=(3.6\pm 1.6)\times 10^{-3} \ ,
 \qquad  y_D=\frac{\Delta \Gamma_D}{2\Gamma_D}=(6.1\pm 0.7)\times 10^{-3} \ ,
  \end{align}
where $\Delta M_D$ and $\Delta \Gamma_D$ are the differences of the masses and the decay widths 
between  the mass eigenstates of the $D$ meson, respectively, 
 and $\Gamma_D$ is the averaged  decay width of the  $D$ meson.
Since the SM prediction of $\Delta M_D$ at the short distance 
is much suppressed compared with the experimental value due to the bottom quark loop,
 the SUSY contribution may be enhanced.

%%%%%%%%%%%%%%%%%%%%%%%%%%%%%%%%%%%%%%%%%%%%%%%%%%%%%%%%%%%
%%%%%%%%%%%%%%%%%%%%%%%%%%%%%%%%%%%%%%%%%%%%%%%%%%%%%%%%%%%
\section{Results and Discussions}
%%%%%%%%%%%%%%%%%%%%%%%%%%%%%%%%%%%%%%%%%%%%%%%%%%%%%%%%%%%
%%%%%%%%%%%%%%%%%%%%%%%%%%%%%%%%%%%%%%%%%%%%%%%%%%%%%%%%%%%
\label{sec:Numerical}
Let us estimate the SUSY contribution of the low energy FCNC. 
We calculate the SUSY mass spectrum at $Q_0=10, 50, 100,1000$ TeV
and interpolate the each mass of the SUSY particle in the region of $Q_0=10$-$1000$ TeV.
This approximation is satisfied within ${\cal O}(10\%)$. Therefore, our numerical results
 should be taken with the ambiguity of  ${\cal O}(10\%)$.
 The mass spectrum at $Q_0=10$ TeV is presented in Appendix A.
See Refs.\cite{Tanimoto:2014eva}, \cite{Tanimoto:2015ota} for the mass spectrum at $Q_0=50$ TeV.

Then, we have four mixing angles $\theta_{13}^{L(R)}$ and  $\theta_{23}^{L(R)}$, 
  five phase $\phi_{13}^{L(R)}$, $\phi_{23}^{L(R)}$,  $\phi$.
 We reduce the number of parameters
by taking  $\sin\theta_{ij}^{\rm L} =\sin \theta_{ij}^{\rm R}\equiv s_{ij}$ for simplicity.
In the numerical calculations,  we scan the phases of Eq. (\ref{mixing})
in the region of $0 \sim 2\pi$ for fixed  $s_{ij}$,
where the Cabibbo angle $0.22$ and the large angle $0.5$ are taken as the typical mixing.
Other relevant input parameters such as quark masses $m_c$, $m_b$,  the CKM parameters
 $V_{us} $, $V_{cb}$, $\bar\rho$, $\bar\eta$  and  $f_B$, $f_K$,  etc. have  been presented 
 in our previous paper
Ref. \cite{Shimizu:2013jia}, which are referred from  UTfit Collaboration \cite{UTfit}
and  PDG \cite{PDG}.

\subsection{$B^0$ and $B_s$ meson systems}

At first, we examine the SUSY contribution in the $\Delta B=2$ process.
We show the SUSY scale $m_{\tilde Q}\equiv Q_0$ dependence of the SUSY contributions of 
$\Delta M_{B^0}$ and  $\Delta M_{B_s}$ in Figure 1(a) and (b),
 where the experimental central value is shown by the red line.
 The experimental error-bars are $1\%$ and $0.1\%$ levels for $\Delta M_{B^0}$ and  $\Delta M_{B_s}$,
 respectively.
 We take $s_{13}=s_{23}=0.22, 0.5$. There is no phase dependence in our predictions.
 It is found that the SUSY contributions in $\Delta M_{B^0}$ and $\Delta M_{B_s}$
are at most $1.5\%$ and $0.1\%$
 at $m_{\tilde Q}=10$ TeV, respectively.
Namely,  the high-scale SUSY  cannot explain 
the NP contributions of $h_d=0.1$-$0.35$ and $h_s=0.15$-$0.25$, which have been discussed
in Eq.(\ref{NP}). 
 As  $m_{\tilde Q}$ increases,
 the SUSY contributions of both $\Delta M_{B^0}$ and $\Delta M_{B_s}$ decrease approximately with
 the power of  $1/m_{\tilde Q}^2$.
 Thus, there is no hope to observe the SUSY contribution in  the $\Delta B=2$ process
 for the high-scale SUSY.
 It should be noted that the SM predictions are comparable to these experimental data.
 
 %%%%%%%%%%%%%  CP violations %%%%%%%%%%%%
 The related phenomena are the CP violation of the non-leptonic decays 
  $B^0\to J/\psi  K_S$ and $B_s\to J/\psi  \phi$.
  The recent experimental data of these phases are
 \cite{Aaij:2013oba,Amhis:2012bh,LHCb:2011aa,LHCb:2011ab}
\begin{equation}
\sin \phi _d=0.679\pm 0.020\ , \qquad \phi_s=0.07\pm 0.09\pm 0.01 \ ,
\label{phasedata}
\end{equation}
 in which the contribution of the gluino-squark-quark interaction may be included.
 The NP contributions in  $\phi_d$ and $\phi_s$ are expressed in terms of 
the  parameters of Eq.(\ref{NP}) as \cite{Shimizu:2013jia}:
\begin{equation}
\phi _d=2\beta_d+\arg(1+h_d e^{2 i\sigma_d}) \ , 
\qquad \phi_s=-2\beta_s+\arg(1+h_s e^{2 i\sigma_s}) \ ,
\label{phiSUSY}
\end{equation}
where $\beta_d(\beta_s)$ is the one angle of the unitarity triangle giving by
 the CKM matrix elements of the SM.
However, $h_d$ and $h_s$ in the high-scale SUSY are much suppressed compared with $h_d=0.1$-$0.35$ and $h_s=0.15$-$0.25$ of Eq.(\ref{NP}),
 one cannot find signals of the high-scale SUSY
in the CP violating decays $B^0\to J/\psi  K_S$ and $B_s\to J/\psi  \phi$.

%%%%%%%%%%%%%%%%%%%%%%%%%%%%%%%%%%%%%%%%%%%%%%%%%%%%%%%%%%%
%%%%%%%%%%%%%%%%%%%%%%%%%%%%%%%%%%%%%%%%%%%%%%%%%%%%%%%%%%
%%%%%%%%%%%%%%%%%%%%%%%%%%%%%%%%%%%%%%%%%%%%%%%%%%%%%%%%%%%
\begin{figure}[t]
\includegraphics[width=8 cm]{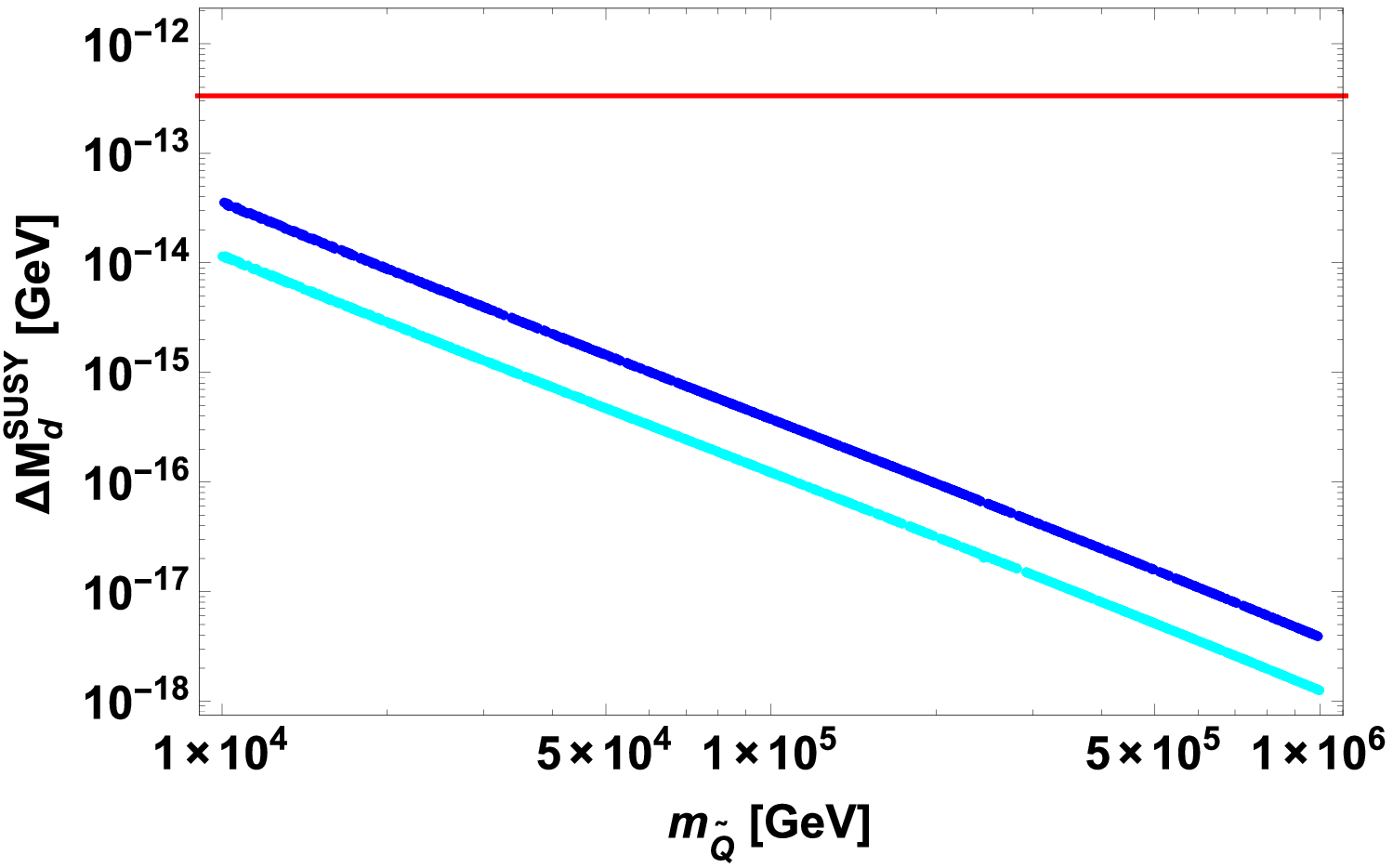}
\hskip 1cm
\includegraphics[width=8 cm]{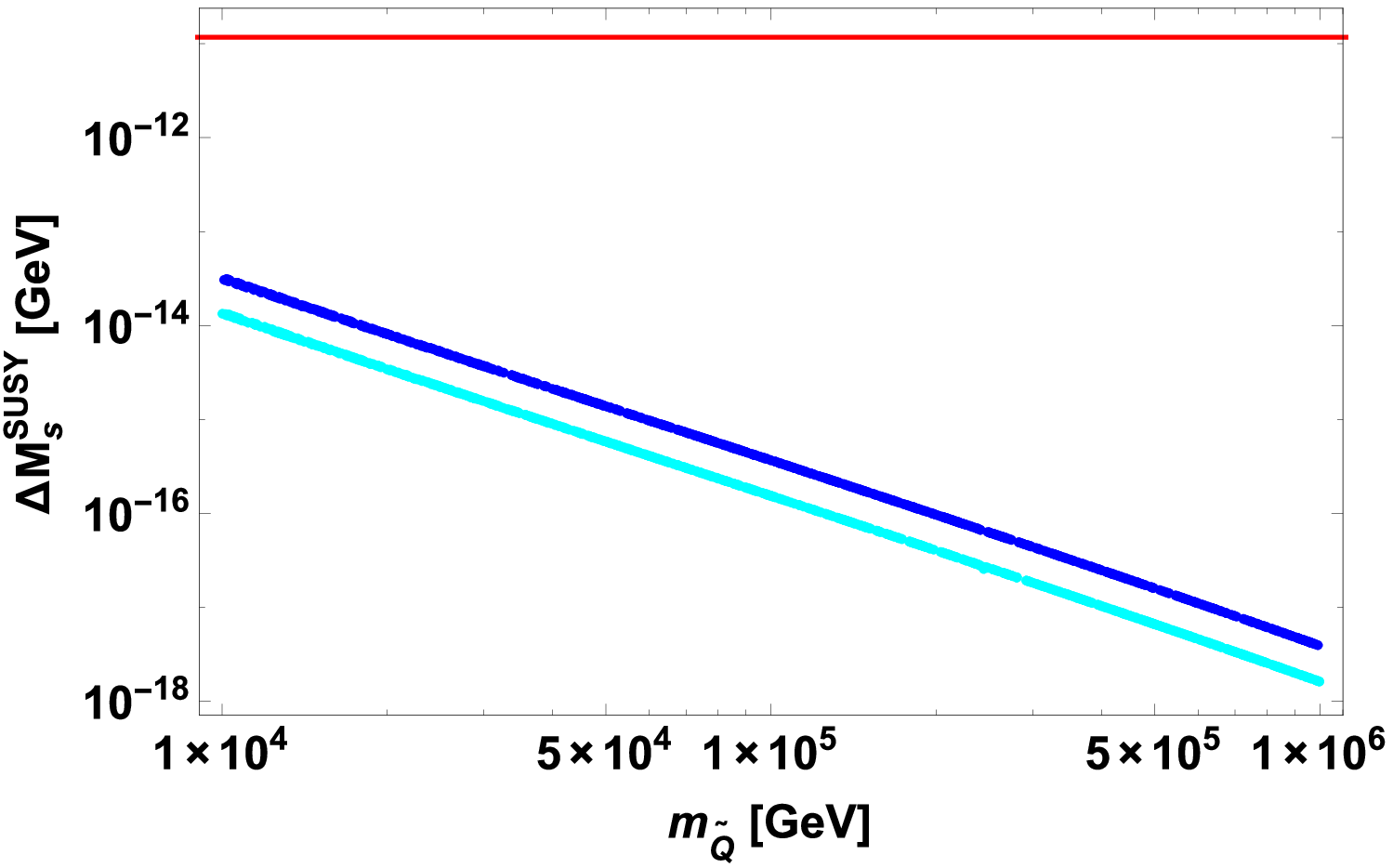}
 \hskip -11.5 cm (a)    \hskip 8.5 cm (b)
\caption{
The SUSY components of  (a) $\Delta M_{B^0}$ and (b) $\Delta M_{B_s}$ versus $m_{\tilde Q}$
for $s_{13}=s_{23}=0.22$ (cyan) and  $0.5$ (blue). 
The horizontal red line denotes the experimental central value.}
\end{figure}
%%%%%%%%%%%%%%%%%%%%%%%%%%%%%%%%%%%%%%%%%%%%%%%%%%%%%%%%%%

\subsection{Neutral $K$ meson system}

At the second step, we examine the neutral $K$ meson.
 We show the SUSY contributions of $\Delta M_{K^0}$ and  $\epsilon_K$ versus 
$m_{\tilde Q}\equiv Q_0$ in Figure 2(a) and (b), 
 where the experimental central value is shown by the horizontal red line.
 The experimental error-bars are $0.2\%$ and $0.5\%$ levels for  $\Delta M_{K^0}$ and  $\epsilon_K$,
 respectively.
 Since $\theta_{12}^{L,R}=0$, the SUSY flavor mixing arise from the second order
  of  $s_{13}\times s_{23}$, where $s_{13}=s_{23}=0.22, 0.5$ are put.

 It is found in Figure 2(a) that the SUSY contribution in $\Delta M_{K^0}$ can be comparable
 to the experimental value in the case of $s_{13}=s_{23}=0.5$ 
 whereas it is suppressed in the case of $s_{13}=s_{23}=0.22$  at $m_{\tilde Q}=10$ TeV.
 Thus, $\Delta M_{K^0}$  constrains  the squark mixing
 of $s_{13}$ and $s_{23}$ around $m_{\tilde Q}=10$ TeV.
 When the SUSY scale increases to more than $20$ TeV, no  SUSY contribution
 is expected.

%%%%%%%%%%%%%%%%%%%%%% epsilon_K %%%%%%%%%%%%%%%%%%%%%%%%%%%
On the other hand,  $\epsilon_K$ is very sensitive to the SUSY contribution
 up to  $100$TeV as seen in Figure 2(b).
 The plot is scattered due to the random  phases of the squark mixing.
 The experimental data of $\epsilon_K$ constrains the squark mixing and phases
 considerably.
 Actually, we have already pointed out that 
the SUSY contribution in $\epsilon_K$ could be $40\%$ and $35\%$ at $m_{\tilde Q}=10, 50$ TeV,
respectively \cite{Tanimoto:2014eva}.  It is  found that this seizable SUSY contribution
 still exist up to $100$ TeV in this work.
 
 In the SM, there is only one CP violating phase.
Therefore, the observed value of  $ \phi_d$ in Eq.(\ref{phiSUSY}),
should be correlated with $\epsilon_K$ in the SM.
According to the recent experimental results, 
it is found that the consistency between the SM prediction and the experimental data of 
$\sin \phi_d$ and  $\epsilon_K$ is marginal.
This fact was pointed out by Buras and Guadagnoli \cite{Buras:2008nn} 
and called as the tension between $\epsilon_K$ and $\sin \phi_d$.
Considering the effect of the SUSY contribution ${\cal O}(10\%)$ in $\epsilon_K$, 
 this tension can be relaxed even if $m_{\tilde Q}=100$ TeV.   
 The precise determination of the unitarity triangle of $B^0$ is required 
in order to find the SUSY contribution of this level.
 
 It is noted that the SUSY contribution of both $\Delta M_{K^0}$ and $\epsilon_K$ also decrease
approximately with the power of $1/m_{\tilde Q}^2$ as $m_{\tilde Q}$ increases up to $1000$ TeV.

%%%%%%%%%%%%%%%%%%%%%%%%%%%%%%%%%%%%%%%%%%%%%%%%%%%%%%%%%%
\begin{figure}[t]
\includegraphics[width=8 cm]{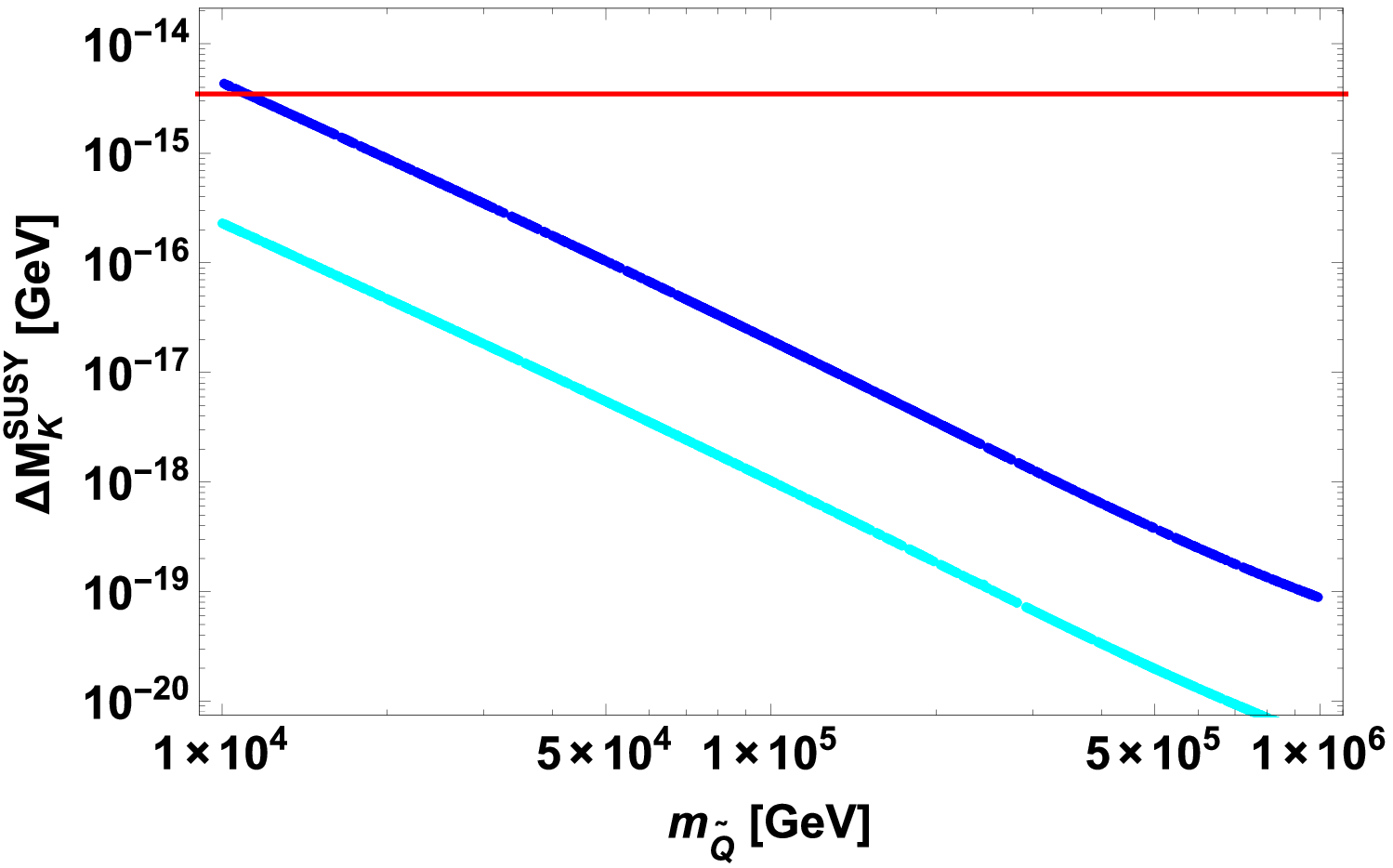}
\hspace{1cm}
\includegraphics[width=8 cm]{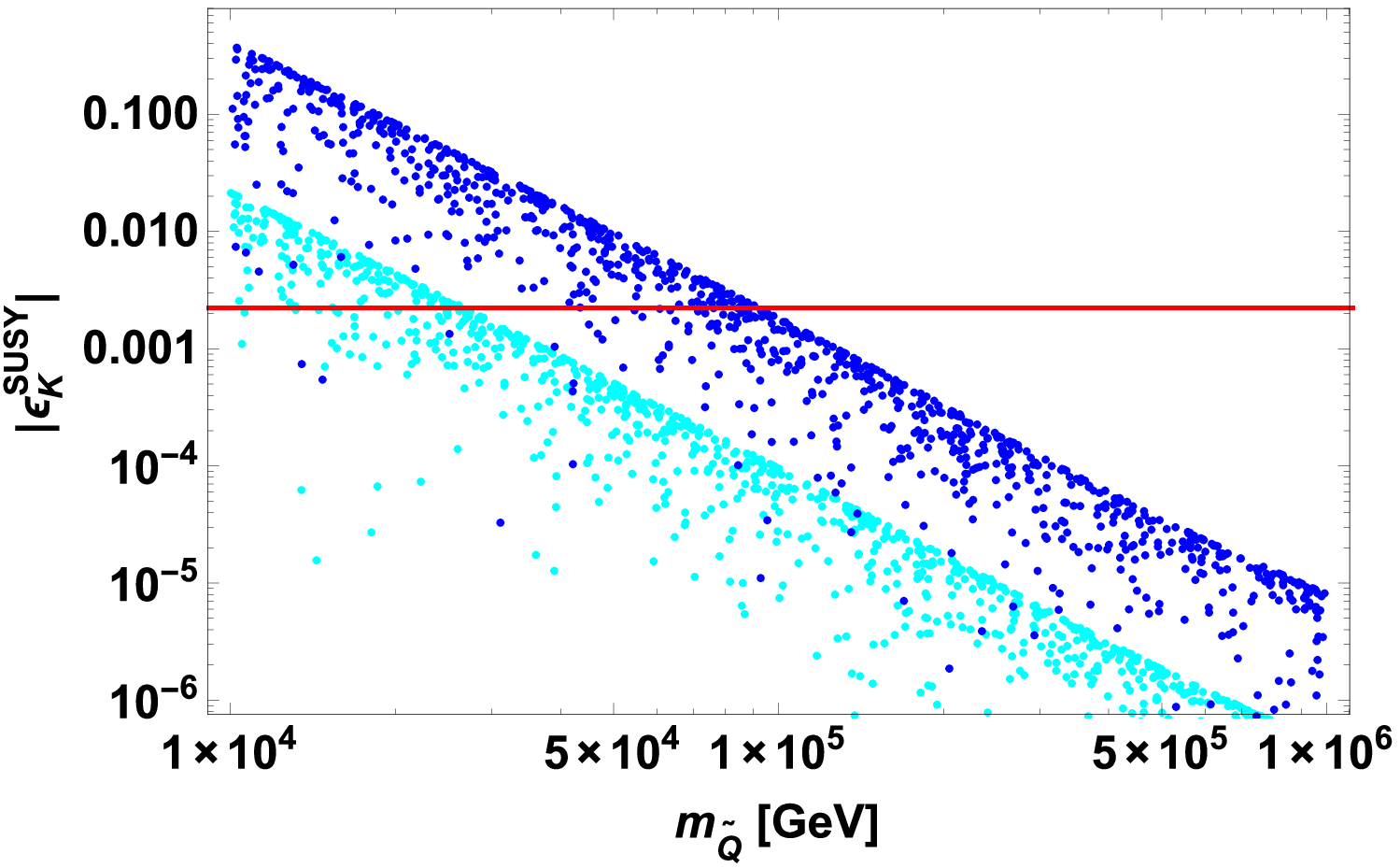}
\hskip -11.5 cm (a)    \hskip 8.5 cm (b)
\caption{The SUSY components of (a) $\Delta M_{K^0}$ and (b) $|\epsilon_K|$ versus $m_{\tilde Q}$
for $s_{13}=s_{23}=0.22$ (cyan) and  $0.5$ (blue).
The horizontal red line denotes the experimental central value.} 
\end{figure}
%%%%%%%%%%%%%%%%%%%%%%%%%%%%%%%%%%%%%%%%%%%%%%%%%%%%%%%%%%%

%%%%%%%%%%%%%%%%%%%%%%%%%%%%%%%%%%%%%%%%%%%%%%%%%%%%%%%%%%
\begin{figure}[t]
\includegraphics[width=8 cm]{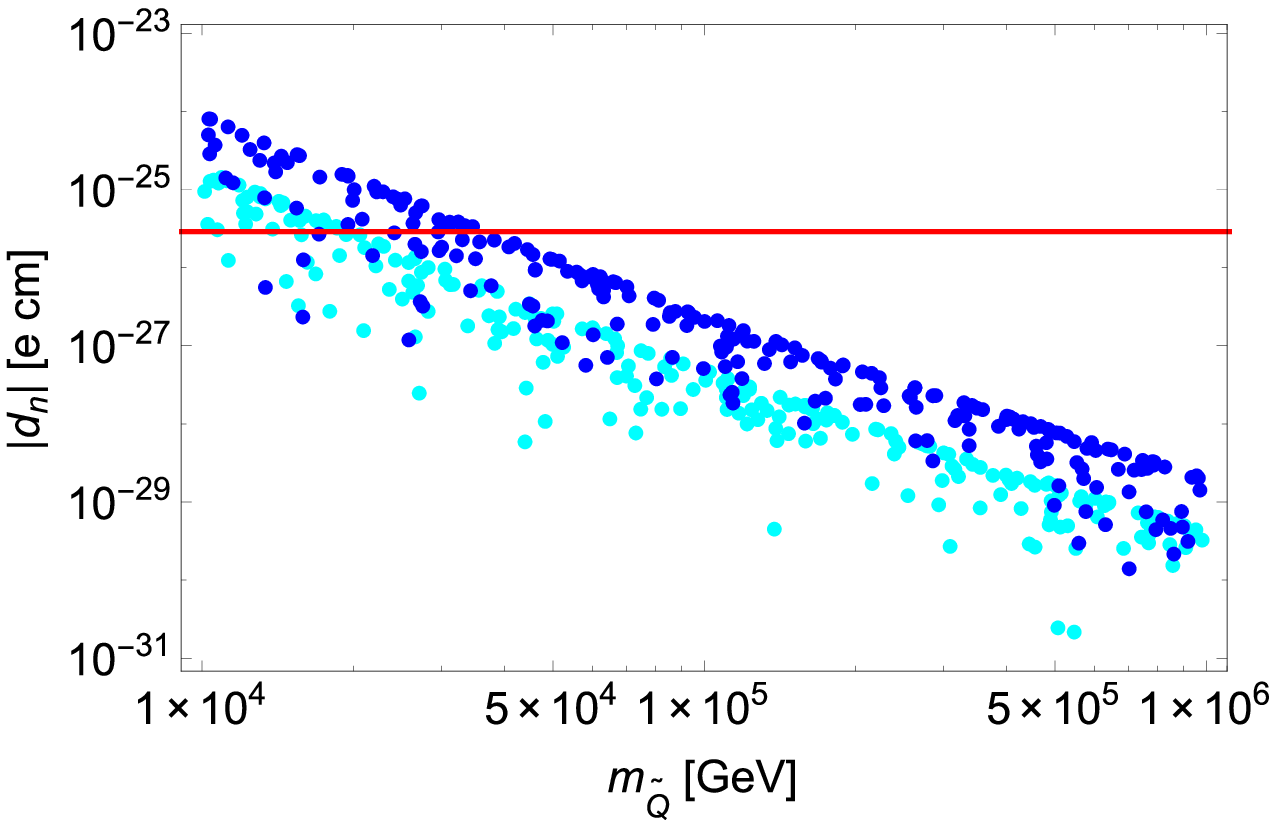}
\hspace{1cm}
\includegraphics[width=8 cm]{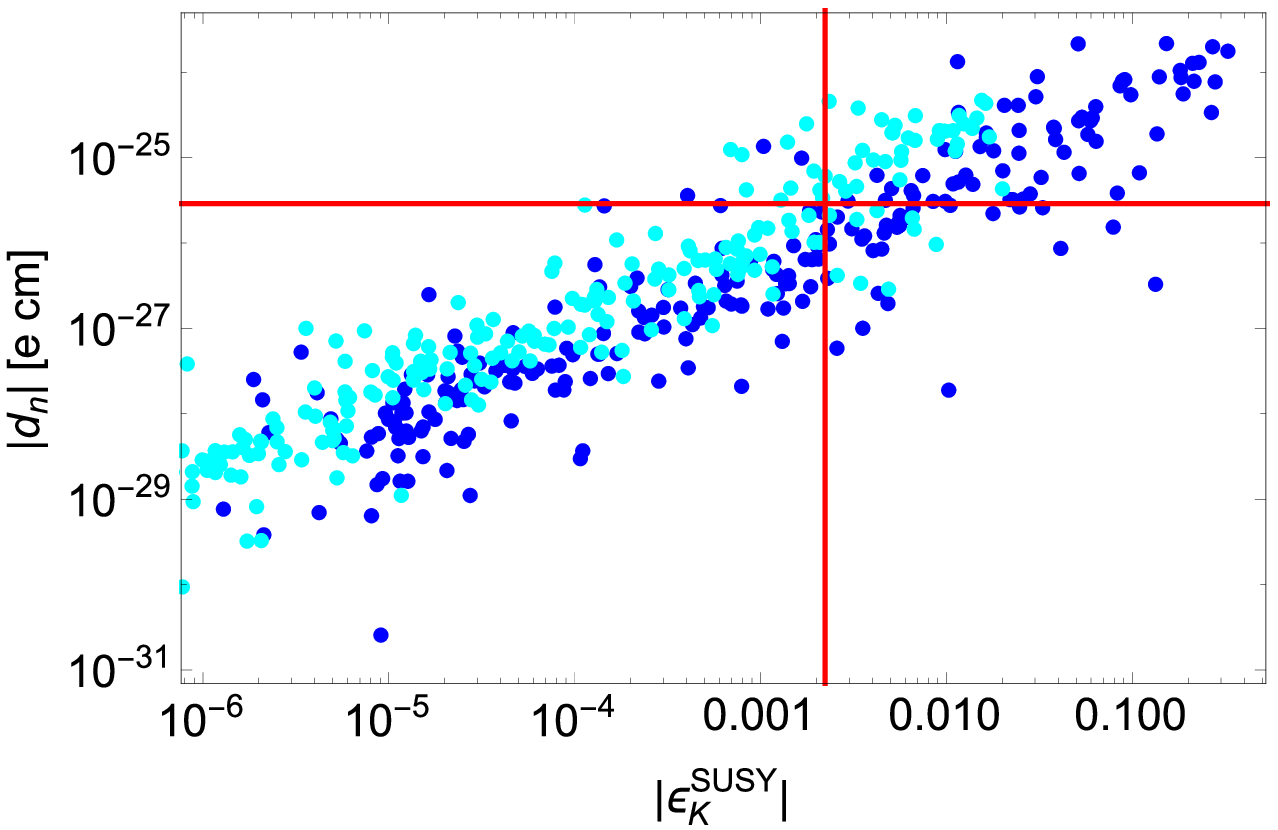}
\hskip -11.5 cm (a)    \hskip 8.5 cm (b)
\caption{The  the neutron EDM versus (a) $m_{\tilde Q}$
and (b) versus   $|\epsilon_K^{\rm SUSY}|$  
for $s_{13}=s_{23}=0.22$ (cyan) and  $0.5$ (blue) for the case of the QCD sum rule.
The horizontal red line denotes the experimental  upper bound of $|d_n|$
and the vertical one is the experimental central value of $|\epsilon_K|$.} 
\end{figure}
%%%%%%%%%%%%%%%%%%%%%%%%%%%%%%%%%%%%%%%%%%%%%%%%%%%%%%%%%%%
%%%%%%%%%%%%%%%%%%%%%%%%%%%%%%%%%%%%%%%%%%%%%%%%%%%%%%%%%%%
\begin{figure}[h!]
\includegraphics[width=8 cm]{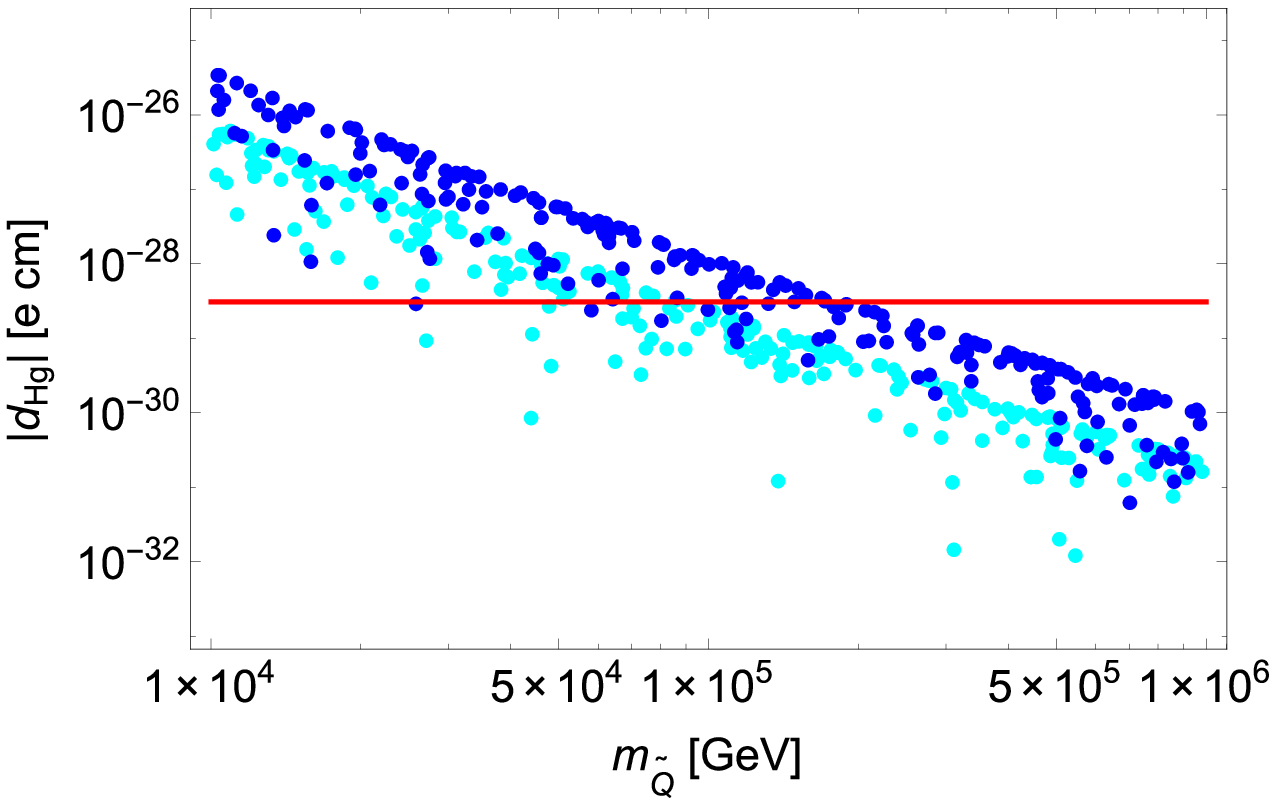}
\hspace{1cm}
\includegraphics[width=8 cm]{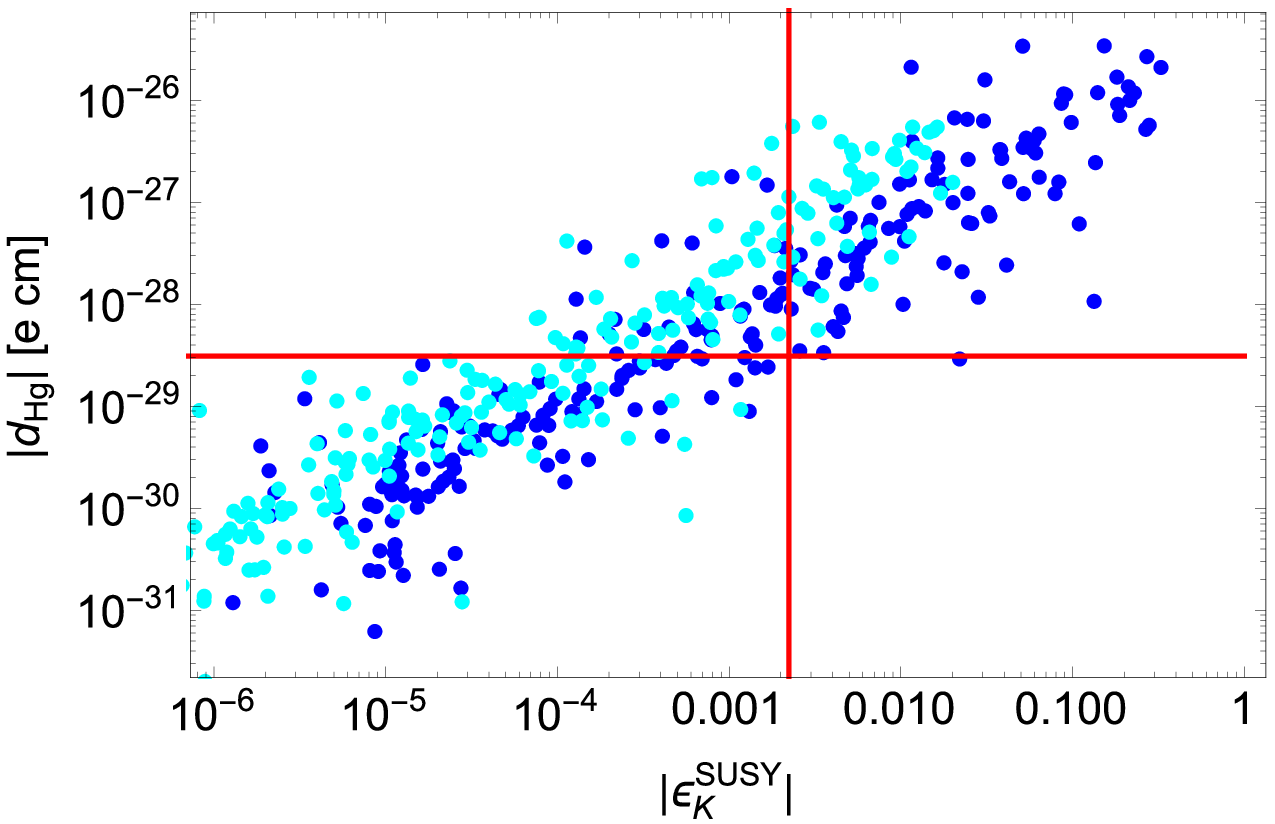}
\hskip -11.5 cm (a)    \hskip 8.5 cm (b)
\caption{The  the mercury EDM versus (a) $m_{\tilde Q}$
and (b) versus   $|\epsilon_K^{\rm SUSY}|$  
for $s_{13}=s_{23}=0.22$ (cyan) and  $0.5$ (blue) for the case of the QCD sum rule.
The horizontal red line denotes the experimental  upper bound of $|d_{Hg}|$
and the vertical one is the experimental central value of $|\epsilon_K|$.} 
\end{figure}
%%%%%%%%%%%%%%%%%%%%%%%%%%%%%%%%%%%%%%%%%%%%%%%%%%%%%%%%%%%

\subsection{The nEDM and HgEDM with $\epsilon_K$}

The nEDM and HgEDM are also sensitive to the SUSY contribution \cite{Fuyuto:2013gla,Chiou:2006qk}.
The gluino-squark interaction leads to the cEDM of quarks,
which give the nEDM as shown in Eqs.(\ref{cedmQCD}) and (\ref{cedmChiral}).
We show the predicted nEDM versus $m_{\tilde Q}$ for the case of the QCD sum rules
 of  Eqs.(\ref{cedmQCD}) in Figure 3(a),
where the upper bound of $|d_n|$ is shown by the red line.
The plot is scattered  due to the random  phases of the squark mixing as well as in the case of
 $\epsilon_K$.
 We find that the contribution of EDM, $d_d$ and $d_u$ occupy around $25\%$ of the neutron EDM.
The SUSY contribution is close to the experimental upper bound up to $50$TeV.
Since  the predicted nEDM depends on the phases of the squark mixing matrix significantly,
we plot the nEDM versus $|\epsilon_K^{\rm SUSY}|$ in Figure 3(b).
It is found that the predicted nEDM is roughly proportional to $|\epsilon_K^{\rm SUSY}|$.
If the SUSY contribution is the level of ${\cal O}(10\%)$ for $\epsilon_K$,
the nEDM is expected to be discovered in the region of $10^{-27}$-$10^{-26}$ecm.
On the other hand, if the nEDM is not observed above $10^{-28}$ecm,
the SUSY contribution of $\epsilon_K$ is below a few $\%$.
Thus, there is the correlation between $d_n$ and $\epsilon_K^{\rm SUSY}$.
%%%%%%%%%%%%%%%%%%%%%%%%%%%%%%%%%%%%%%%%%%%%%%%%%%

We also show the predicted HgEDM versus $m_{\tilde Q}$ for the case of the QCD sum rules
 of  Eq.(\ref{cedmHgQCD}) in Figure 4(a),
where the upper bound of $|d_{Hg}|$ is shown by the red line.
The SUSY contribution is close to the experimental upper bound up to $200$TeV,
which is much higher than the one of the  nEDM.
In Figure 4(b), 
we plot the HgEDM versus $|\epsilon_K^{\rm SUSY}|$.
It is found that the experimental upper bound of the  HgEDM  excludes
completely  $|\epsilon_K^{\rm SUSY}|$ which is inconsistent with the experimental data.
If the SUSY contribution is the level of ${\cal O}(10\%)$ for $\epsilon_K$,
the nEDM is expected to be discovered in the region of $10^{-27}$-$10^{-26}$ecm.
If the HgEDM is not observed above $10^{-29}$ecm,
the SUSY contribution of $\epsilon_K$ is below a few $\%$.
Thus, the mercury EDM gives more significant information for the  gluino-squark interaction
compared with the neutron EDM.
%%%%%%%%%%%%%%%%%%%%%%%%%%%%%%%%%%%%%%%%%%%%%%%%%%%%%

However,
these correlations strongly depend on the
assumptions of  $\theta_{23}^L=\theta_{13}^L$ and  $\theta_{ij}^L=\theta_{ij}^R$.
The deviation from these relations destroys these correlations.
For instance, for the case of
 $\theta_{23}^L\gg \theta_{13}^L$ with  $\theta_{ij}^L=\theta_{ij}^R$,
$\epsilon_K^{\rm SUSY}$ is much suppressed whereas the nEDM and HgEDM are still sizable.
On the other hand, if $\theta_{ij}^L\gg\theta_{ij}^R$ or  $\theta_{ij}^L\gg\theta_{ij}^R$
is realized,  the cEDMs are suppressed because they require the chirality flipping.
In conclusion, the careful studies of the mixing angle relations are required
 to test the correlations between EDMs and $\epsilon_K^{\rm SUSY}$.

%%%%%%%%%%%%%%%%%%%%%%%%%%%%%%%%%%
We should comment on the hadronic model dependence of our numerical result.
For both nEDM and HgEDM, we show the numerical result by using the hadronic model
of the QCD sum rules in Eqs.(\ref{cedmQCD}) and (\ref{cedmHgQCD}).
We have also calculated the EDMs by using the hadronic model
of the chiral perturbation theory in Eqs.(\ref{cedmChiral}) and (\ref{cedmHgChiral}).
For the neutron EDM, 
 the prediction of the chiral perturbation theory  is larger than the one of the QCD sum rule
 at most of  factor two.
However, for the mercury EDM,
 the prediction of  the QCD sum rule  is more than three times larger compared with   the one of the  chiral perturbation theory.
 Thus, predicted EDMs have the ambiguity with the factor $2-3$ from the hadronic model.
%%%%%%%%%%%%%%%%%%%%%%%%%%%%%%%%%%%%%%%%%%

\subsection{$D$-$\bar D$ mixing}

Since the SM prediction of $\Delta M_D$ at the short distance 
is ${\cal O}(10^{-18})$ GeV, 
 which is very small compared with the experimental value due to the bottom quark loop,
 it is important to estimate the SUSY contribution of $\Delta M_D$.
 The mixing angle $\theta_{ij}^{L(R)}$ also appears in the up-type squark mixing matrix
 whereas the down-type squark mixing matrix  contributes to $K^0$, $B^0$ and $B_s$ meson
 systems induced by the gluino-squark-quark interaction.
 
 We show the SUSY component of $\Delta M_{D}$ and $x_D$ versus $m_{\tilde Q}$
for $s_{13}=s_{23}=0.22, 0.5$ in Figure 5. 
At the SUSY scale of $10$ TeV, the SUSY component may be comparable to the observed value.
Although the accurate estimate of the long-distance effect is difficult,
 Cheng and Chiang estimated $x_D$ of order $10^{-3}$ 
from the two body hadronic modes \cite{Cheng:2010rv}. 
This obtained value is consistent with the experimental one.
Therefore,  we should take into account the long-distance effect properly
in order to constrain the SUSY contribution from $\Delta M_D$.
 
%%%%%%%%%%%%%%%%%%%%%%%%%%%%%%%%%%%%%%%%%%%%%%%%%%%%%%%%%%% 
%%%%%%%%%%%%%%%%%%%%%%%%%%%%%%%%%%%%%%%%%%%%%%%%%%%%%%%%%%%
Before closing the presentation of the numerical results, we add a comment on the other gaugino  contribution.
There are additional contributions to the FCNC
induced by chargino  exchanging diagrams.
The chargino contribution to the gluino one is approximately $10\%$ in the above numerical study
of $\Delta F=2$. Thus, the  chargino contributions are the sub-leading ones.

%%%%%%%%%%%%%%%%%%%%%%%%%%%%%%%%%%%%%%%%%%%%%%%%%%%%%%%%%%%
\begin{figure}[]
\includegraphics[width=8 cm]{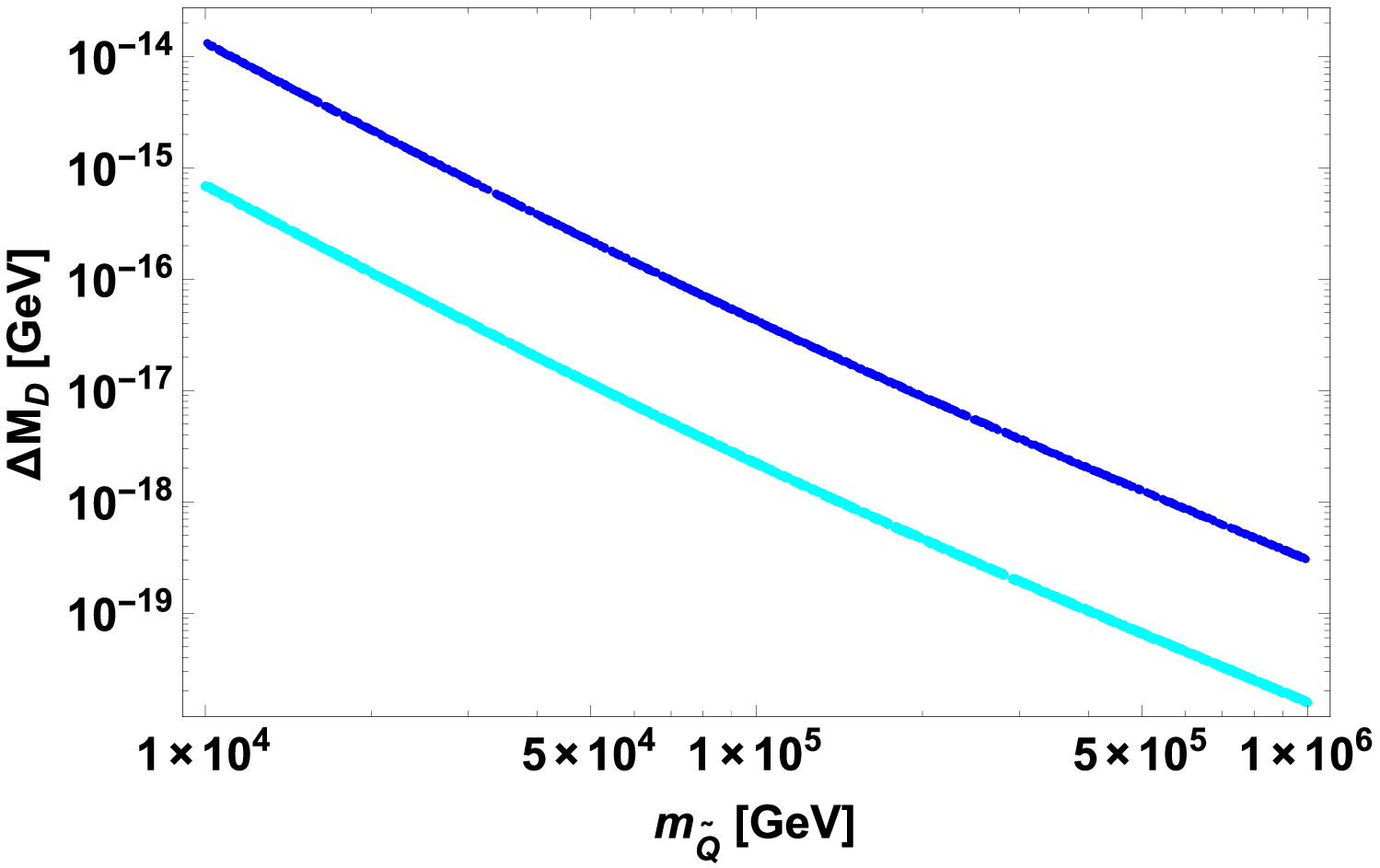}
\hspace{1cm}
\includegraphics[width=8 cm]{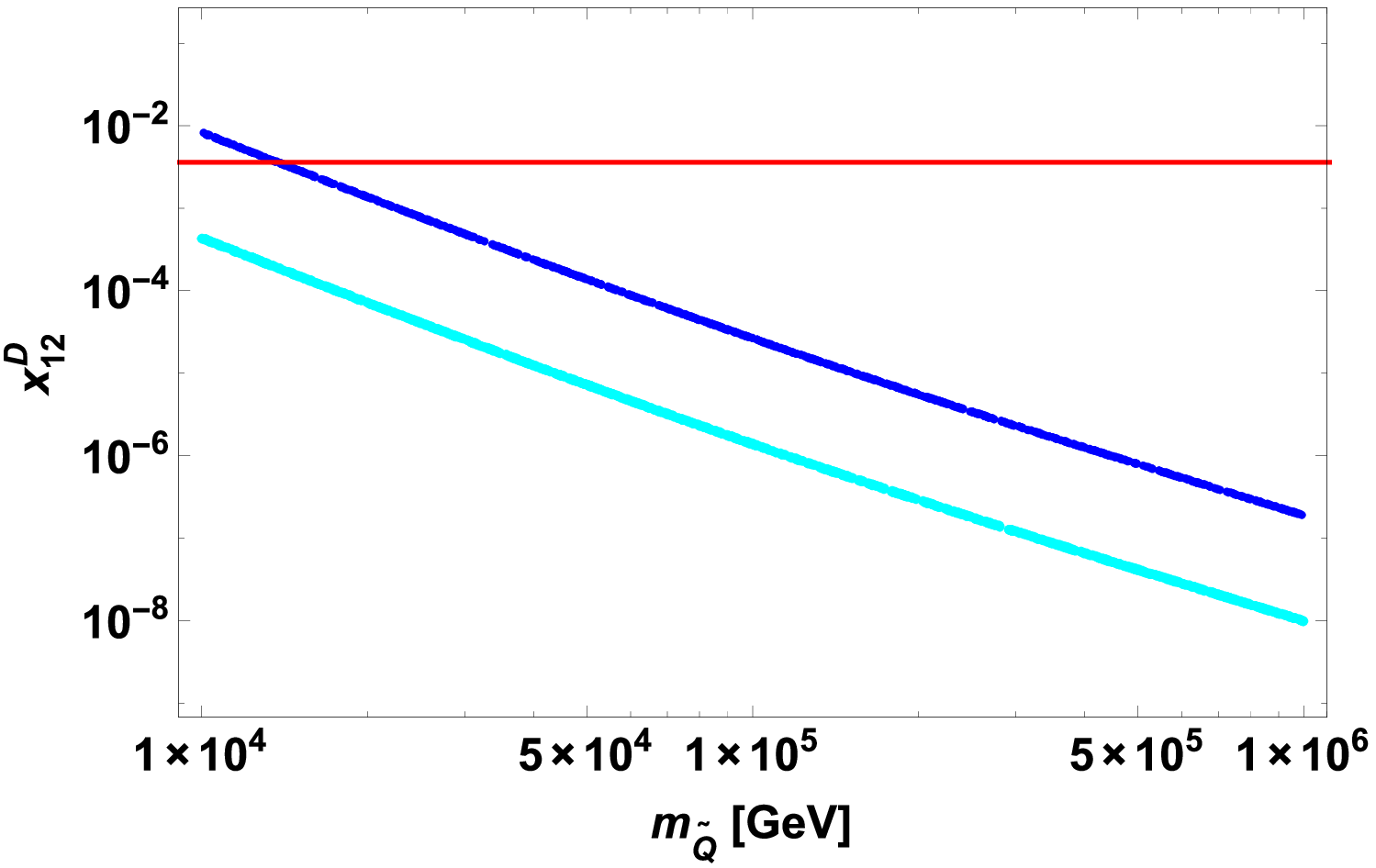}
\hskip -11.5 cm (a)    \hskip 8.5 cm (b)
\caption{The SUSY component of (a)$\Delta M_{D}$ and (b)$x_D$ versus $m_{\tilde Q}$
for $s_{13}=s_{23}=0.22$ (cyan) and  $0.5$ (blue). 
The horizontal red line denotes the experimental central value.
} 
\end{figure}
%%%%%%%%%%%%%%%%%%%%%%%%%%%%%%%%%%%%%%%%%%%%%%%%%%%%%%%%%%%
%%%%%%%%%%%%%%%%%%%%%%%%%%%%%%%%%%%%%%%%%%%%%%%%%%%%%%%%%%%

%%%%%%%%%%%%%%%%%%%%%%%%%%%%%%%%%%%%%%%%%%%%%%%%%%%%%%%%%%%
%%%%%%%%%%%%%%%%%%%%%%%%%%%%%%%%%%%%%%%%%%%%%%%%%%%%%%%%%%%
\section{Summary}
%%%%%%%%%%%%%%%%%%%%%%%%%%%%%%%%%%%%%%%%%%%%%%%%%%%%%%%%%%%
%%%%%%%%%%%%%%%%%%%%%%%%%%%%%%%%%%%%%%%%%%%%%%%%%%%%%%%%%%%

 We discussed the sensitivity of the high-scale SUSY at $10$-$1000$ TeV in the 
  $B^0$, $B_s$ and $K^0$ meson systems. Furthermore, we have  also discussed the sensitivity to the $D$-$\bar D$ mixing, the neutron EDM and the mercury EDM.
  In order to estimate the contribution of the squark flavor mixing to these FCNC,
we calculate the squark mass spectrum, which is consistent with
   the recent Higgs discovery. 
   
   The SUSY contributions in $\Delta M_{B^0}$ and $\Delta M_{B_s}$
are at most $1.5\%$ and $0.1\%$
 at $m_{\tilde Q}=10$ TeV, respectively. 
 As  $m_{\tilde Q}$ increases,
 the SUSY contributions of both $\Delta M_{B^0}$ and $\Delta M_{B_s}$ decrease approximately with
 the power of  $1/m_{\tilde Q}^2$.
  Therefore, the SUSY scale increases to more than $10$ TeV, no signal of the SUSY
 is expected. 
 On the other hand,
 the SUSY contribution in $\Delta M_{K^0}$ can be comparable
 to the experimental value in the case of $s_{13}=s_{23}=0.5$ 
 whereas it is suppressed in the case of $s_{13}=s_{23}=0.22$  at $m_{\tilde Q}=10$ TeV.
Furthermore,
the SUSY contribution in $\epsilon_K$ could be large, around $40\%$ in the region
of  the SUSY scale   $10$-$100$ TeV.
By considering the effect of the SUSY contribution ${\cal O}(10\%)$ in $\epsilon_K$, 
 the tension between $\epsilon_K$ and $\sin \phi_d$
can be relaxed even if the SUSY scale is  $100$ TeV.

 The neutron EDM and   the mercury EDM are also sensitive to the SUSY contribution induced by
the gluino-squark interaction.
The $|d_n|$ is expected to be 
close to the experimental upper bound even if the SUSY scale is $50$ TeV. 
The predicted nEDM is roughly proportional to $|\epsilon_K^{\rm SUSY}|$.
If the SUSY contribution is the level of ${\cal O}(10\%)$ for $\epsilon_K$,
the  $|d_n|$ is expected to be discovered in the region of $10^{-27}$-$10^{-26}$cm.
For the  $|d_{Hg}|$, 
the SUSY contribution is close to the experimental upper bound up to $200$TeV,
which is much higher than the one of the  nEDM.
If the HgEDM is not observed above $10^{-29}$cm,
the SUSY contribution of $\epsilon_K$ is below a few $\%$.
Thus, the mercury EDM gives more significant information for the  gluino-squark interaction
compared with the neutron EDM.
It may be important to give a comment that these predictions depend strongly  on the
assumptions of  $\theta_{23}^L=\theta_{13}^L$ and  $\theta_{ij}^L=\theta_{ij}^R$.
The deviation from these relations destroys these correlations.
In conclusion, the careful studies of the mixing angle relations are required
 to test the correlations between EDMs and $\epsilon_K^{\rm SUSY}$.
 The  predicted EDMs have also the ambiguity with the factor $2-3$ from the hadronic model.

Since the SM prediction of $\Delta M_D$ at the short distance 
is ${\cal O}(10^{-18})$ GeV, 
 which is very small compared with the experimental value,
 it is important to estimate the SUSY contribution of $\Delta M_D$.
 
 In conclusion,  the more detailed studies of  
$K^0$ meson system, the  EDMs of the neutron and  mercury are required 
 in order to probe the high-scale SUSY at  $10$-$1000$ TeV.

%%%%% acknowledgement %%%%%
\vspace{0.5 cm}
\noindent
{\bf Acknowledgment}

\vskip 0.3 cm
This work is supported by JSPS Grand-in-Aid for Scientific Research,
  24654062 and  25-5222.

%%%%%%%%%%%%%%%%%%%%%%%%%%%%%%%%%%%%%%%%%%%%%%%%%%%%%%%%%%%%
%%%%%%%%%%%%%%%%%%%%%%%  Appendices %%%%%%%%%%%%%%%%%%%%%%%%
%%%%%%%%%%%%%%%%%%%%%%%%%%%%%%%%%%%%%%%%%%%%%%%%%%%%%%%%%%%%
\newpage
\appendix{}
\section*{Appendix A : Running of  SUSY particle masses}

In the framework of the MSSM,
one  obtains the SUSY particle spectrum which is consistent with the observed Higgs mass.
The numerical analyses have been given in Refs.
\cite{Delgado:2013gza, Giudice:2006sn}.
At the SUSY breaking scale $\Lambda$, the quadratic terms in the MSSM potential is given as
\begin{equation}
 V_2=m_1^2 |H_1|^2+ m_2^2 |H_2|^2+m_3^2 (H_1\cdot H_2+ h.c.) \ .
\end{equation}
The mass eigenvalues at  the $H_1$ and $\tilde H_2\equiv \epsilon H_2^*$ system are given
\begin{equation}
 m_\mp^2 = \frac{m_1^2+m_2^2}{2}\mp \sqrt{\left (\frac{m_1^2-m_2^2}{2}\right )^2+m_3^4} \ .
\end{equation}
Suppose that the MSSM matches with the SM at the SUSY mass scale $Q_0\equiv m_0$.
Then, the smaller one $m_{-}^2$ is identified to be the mass squared of the SM Higgs $H$ with the tachyonic mass.
The larger one $m_{+}^2$ is  the mass squared of the orthogonal combination ${\cal H}$, which
is decoupled from the SM at $Q_0$, that is, $m_{{\cal H}}\simeq Q_0$. Therefore, we have
\begin{eqnarray}
 m_{-}^2=-m^2(Q_0)\ , \qquad  m_{+}^2=m_{\cal H}^2(Q_0) =m_1^2+m_2^2+m^2 \ ,
\end{eqnarray}
with 
\begin{eqnarray}
 m_{3}^4=(m_1^2+m^2)(m_2^2+m^2) \ ,
\end{eqnarray}
which leads to the mixing angle between $H_1$ and $\tilde H_2$, $\beta$ as follows:
\begin{eqnarray}
 \tan^2\beta=\frac{m_1^2+m^2}{m_2^2+m^2} \ , \qquad 
 H=  \cos\beta H_1 +\sin\beta \tilde H_2  \ , \qquad
 {\cal H} =-\sin\beta H_1 +\cos\beta \tilde H_2  \ .
\end{eqnarray}
Thus,  the Higgs mass parameter $m^2$ is expressed in terms of $m_1^2$, $m_2^2$ and $\tan\beta$:
\begin{eqnarray}
m^2=\frac{m_1^2-m_2^2\tan^2\beta}{\tan^2\beta -1} \ .
\end{eqnarray}
Below the  $Q_0$ scale, in which the SM emerges, the scalar potential is  the SM one as follows:
\begin{equation}
 V_{SM}=-m^2 |H|^2+\frac{\lambda}{2} |H|^4 \ .
\end{equation}
Here, the Higgs coupling $\lambda$ is given in terms of the SUSY parameters
at the leading order as
 \begin{equation}
 \lambda(Q_0)=\frac{1}{4} (g^2+g'^2)\cos^2 2\beta +\frac{3h_t^2}{8\pi^2} X_t^2 \left (1- \frac{X_t^2}{12}\right ) \ , \qquad
 X_t=\frac{A_t(Q_0)-\mu(Q_0)\cot\beta}{ Q_0 } \ ,
\end{equation}
and $h_t$ is the top Yukawa coupling of the SM.
 The parameters $m_2$ and $\lambda$ run with the SM Renormalization Group Equation down to the electroweak scale $Q_{EW}=m_H$, and then give
\begin{equation}
 m_H^2=2m^2(m_H)=\lambda(m_H)v^2\ .
\end{equation}
It is easily seen  that the VEV of Higgs, $\langle H \rangle $ is $v$,  and $\langle {\cal H}\rangle=0$,
taking account of $\langle H_1\rangle =v\cos\beta$ and $\langle H_2\rangle =v\sin\beta$,
 where $v=246$ GeV.

Let us  fix $m_H=125$ GeV, which gives $\lambda(Q_0)$ and $m^2(Q_0)$. This experimental input constrains the SUSY mass spectrum of the MSSM.
We consider the some universal soft breaking parameters at the SUSY breaking scale $\Lambda$ as follows:
\begin{eqnarray}
&&m_{\tilde Q_i}(\Lambda)=m_{\tilde U^c_i}(\Lambda)=m_{\tilde D^c_i}(\Lambda)=
m_{\tilde L_i}(\Lambda)=m_{\tilde E^c_i}(\Lambda)=m_0^2 \ (i=1,2,3) \ , \nonumber \\
&&M_1(\Lambda)=M_2(\Lambda)=M_3(\Lambda)=m_{1/2}  \ , \qquad 
m_1^2({\Lambda})= m_2^2({\Lambda})=m_0^2 \ , \nonumber \\
&&A_U({\Lambda})=A_0 y_U(\Lambda)\ , \quad A_D({\Lambda})=A_0 y_D(\Lambda)\ , 
\quad A_E({\Lambda})=A_0 y_E(\Lambda)\ .
\end{eqnarray}
Therefore, there is no flavor mixing at $\Lambda$ in the MSSM.  
However, in order to consider the non-minimal flavor mixing framework, we allow the off diagonal components of the squark mass matrices at the $10\%$ level, which leads to the
flavor mixing of order $0.1$.
 We take these flavor mixing angles as free parameters  at low energies.

Now, we have the SUSY five parameters, $\Lambda$, $\tan\beta$, $m_0$, $m_{1/2}$, $A_0$,
where $Q_0=m_0$. In addition to these parameters, we take $\mu=Q_0$.
Inputing  $m_H=125$ GeV and taking $m_{{\cal H}}\simeq Q_0$, we can obtain the SUSY spectrum
 for the fixed $Q_0$ and $\tan\beta$.
 
We present the SUSY mass spectrum at $Q_0=10$ TeV.
The input parameter set and the obtained SUSY mass spectra at $Q_0$ are  summarized in Table 1,
where we use $\overline m_t (m_t)=163.5\pm 2$ GeV \cite{PDG,UTfit}.
 These parameter sets are easily found  from the  work  in Ref.\cite{Delgado:2013gza}.
\vskip 0.5 cm
\begin{table}[h!]
\begin{center}
\begin{tabular}{|l|l|}\hline
 \quad Input at $\Lambda$ and $Q_0$ & \qquad  \qquad \qquad \quad Output at $Q_0$ \\
\hline
  & \\
 at $\Lambda=10^{17}$ GeV,  & $m_{\tilde g}=12.8$ TeV,  \ $m_{\tilde W}=5.2$ TeV,  
\ $m_{\tilde B}=2.9$ TeV\\ 
 \quad $m_0=10$ TeV,  & $m_{\tilde b_L}=m_{\tilde t_L}=12.2$ TeV  \\
 \quad $m_{1/2}=6.2$ TeV, &  $m_{\tilde b_R}=14.1$ TeV, \  $m_{\tilde t_R}=8.4$ TeV\\
 \quad  $A_0=25.803$ TeV; & $m_{\tilde s_L , \tilde d_L}=m_{\tilde c_L , \tilde u_L}=15.1$ TeV \\
 at $Q_0=10$ TeV, &   $m_{\tilde s_R,\tilde d_R}\simeq m_{\tilde c_R,\tilde u_R}=14.6$ TeV,  
\ $m_{{\cal H}}=13.7$ TeV  \\
\quad  $\mu=10$ TeV,&  $m_{\tilde \tau_{L}}=m_{\tilde {\nu_\tau}_L}=10.4$ TeV, \  
$m_{\tilde \tau_R}=9.3$ TeV 
\\
 \quad $\tan\beta=10$&  
 $m_{\tilde \mu_L, \tilde e_{L}}=m_{\tilde {\nu_\mu}_L, \tilde {\nu_e}_L}=10.8$ TeV, 
 \ $m_{\tilde \mu_R, \tilde e_R}=10.3$ TeV \\
  &  $X_t=-0.22$, \quad  $\lambda_H=0.126$ 
\\
\hline
\end{tabular}
\end{center}
\caption{Input parameters at $\Lambda$ and  the obtained SUSY spectra at
$Q_0=10$ TeV.}
\label{tab:inputparameters}
\end{table}
As seen in Table 1, the first and second family squarks are degenerate in their masses, on the other hand,
the third ones  split due to the large RGE's effect.
Therefore, the mixing angle between the first and second family squarks vanishes, but the 
mixing angles between the first-third and the second-third family squarks are produced at the $Q_0$ scale.
The left-right mixing angle  between   $\tilde b_{L}$ and $\tilde b_{R}$ is given as 
\begin{eqnarray}
\theta\simeq \frac{m_b (A_b(Q_0)-\mu\tan\beta)}{m_{\tilde b_{L}}^2-m_{\tilde b_{R}}^2} \ ,
\label{leftrightmixing}
\end{eqnarray}
which is very small, ${\cal O}(0.01)$ at $10$ TeV.
The lightest squark is the right-handed stop and 
 the lightest gaugino is the Bino.

%%%%%%%%%%%%%%%%%%%%%%%%%%%%%%%%%%%%%%%%%%%%%%%%%
\section*{Appendix B : Squark contribution in $\Delta F=2$ process}

The $\Delta F=2$ effective Lagrangian from the gluino-sbottom-quark interaction  is given as
\cite{GotoNote}:
\begin{align}
\mathcal{L}_{\text{eff}}^{\Delta F=2}=-\frac{1}{2}\left [C_{VLL}O_{VLL}+C_{VRR}O_{VRR}\right ] 
-\frac{1}{2}\sum _{i=1}^2
\left [C_{SLL}^{(i)}O_{SLL}^{(i)}+C_{SRR}^{(i)}O_{SRR}^{(i)}+C_{SLR}^{(i)}O_{SLR}^{(i)}\right ],
\label{Lagrangian-DeltaF=2}
\end{align}
where
\begin{eqnarray}
&&O_{VLL}=(\bar q_a \gamma^\mu L Q^a)(\bar q_b \gamma^\mu L Q^b), \qquad
O_{VRR}=(\bar q_a \gamma^\mu R Q^a)(\bar q_b \gamma^\mu R Q^b), \nonumber \\
&&O_{SLL}^{(1)}=(\bar q_a L Q^a)(\bar q_b  L Q^b), \qquad
O_{SLL}^{(2)}=(\bar q_a L Q^b)(\bar q_b  L Q^a), \nonumber \\
&&O_{SRR}^{(1)}=(\bar q_a R Q^a)(\bar q_b  R Q^b), \qquad
O_{SRR}^{(2)}=(\bar q_a R Q^b)(\bar q_b  R Q^a), \nonumber \\
&&O_{SLR}^{(1)}=(\bar q_a L Q^a)(\bar q_b  R Q^b), \qquad
O_{SLR}^{(2)}=(\bar q_a L Q^b)(\bar q_b  R Q^a), 
\end{eqnarray}
with   $(P,Q,q)=(B^0,b,d),~(B_s,b,s),~(K^0,s,d)$.
The  $L,R$ denote $(1\pm \gamma_5)/2$, and $a,b$ are color indices.  
Then, the $P^0$-$\bar P^0$ mixing, $M_{12}$, is written as: 
\begin{equation}
M_{12}=-\frac{1}{2m_P}\langle P^0|\mathcal{L}_{\text{eff}}^{\Delta F=2}|\bar P^0\rangle \ .
\end{equation}
The hadronic matrix elements are given in terms of the non-perturbative
parameters  $B_i$ as: 
\begin{align}
\langle P^0|\mathcal{O}_{VLL}|\bar P^0\rangle &=\frac{2}{3}m_P^2f_P^2B_1, \quad 
\langle P^0|\mathcal{O}_{VRR}|\bar P^0\rangle =\langle P^0|\mathcal{O}_{VLL}|\bar P^0\rangle ,\nonumber \\
\langle P^0|\mathcal{O}_{SLL}^{(1)}|\bar P^0\rangle &=-\frac{5}{12}m_P^2f_P^2R_PB_2, \quad 
\langle P^0|\mathcal{O}_{SRR}^{(1)}|\bar P^0\rangle =\langle P^0|\mathcal{O}_{SLL}^{(1)}|\bar P^0\rangle ,\nonumber \\
\langle P^0|\mathcal{O}_{SLL}^{(2)}|\bar P^0\rangle &=\frac{1}{12}m_P^2f_P^2R_PB_3, \quad 
\langle P^0|\mathcal{O}_{SRR}^{(2)}|\bar P^0\rangle =\langle P^0|\mathcal{O}_{SLL}^{(2)}|\bar P^0\rangle ,\nonumber \\
\langle P^0|\mathcal{O}_{SLR}^{(1)}|\bar P^0\rangle &=\frac{1}{2}m_P^2f_P^2R_PB_4, \quad 
\langle P^0|\mathcal{O}_{SLR}^{(2)}|\bar P^0\rangle =\frac{1}{6}m_P^2f_P^2R_PB_5,
\end{align}
where 
\begin{equation}
R_P=\left (\frac{m_P}{m_Q+m_q}\right )^2.
\end{equation}

The Wilson coefficients for the gluino contribution in Eq.~(\ref{Lagrangian-DeltaF=2}) are written as \cite{GotoNote}:

\begin{align}
C_{VLL}(m_{\tilde g})&=\frac{\alpha _s^2}{m_{\tilde g}^2}\sum _{I,J=1}^6
(\lambda _{GLL}^{(d)})_I^{ij}(\lambda _{GLL}^{(d)})_J^{ij}
\left [\frac{11}{18}g_{2[1]}(x_I^{\tilde g},x_J^{\tilde g})
+\frac{2}{9}g_{1[1]}(x_I^{\tilde g},x_J^{\tilde g})\right ],\nonumber \\
C_{VRR}(m_{\tilde g})&=C_{VLL}(m_{\tilde g})(L\leftrightarrow R),\nonumber \\
C_{SRR}^{(1)}(m_{\tilde g})&=\frac{\alpha _s^2}{m_{\tilde g}^2}\sum _{I,J=1}^6
(\lambda _{GLR}^{(d)})_I^{ij}(\lambda _{GLR}^{(d)})_J^{ij}
\frac{17}{9}g_{1[1]}(x_I^{\tilde g},x_J^{\tilde g}),\nonumber \\
C_{SLL}^{(1)}(m_{\tilde g})&=C_{SRR}^{(1)}(m_{\tilde g})(L\leftrightarrow R),\nonumber \\
C_{SRR}^{(2)}(m_{\tilde g})&=\frac{\alpha _s^2}{m_{\tilde g}^2}\sum _{I,J=1}^6
(\lambda _{GLR}^{(d)})_I^{ij}(\lambda _{GLR}^{(d)})_J^{ij}
\left (-\frac{1}{3}\right )g_{1[1]}(x_I^{\tilde g},x_J^{\tilde g}),\nonumber \\
C_{SLL}^{(2)}(m_{\tilde g})&=C_{SRR}^{(2)}(m_{\tilde g})(L\leftrightarrow R),\nonumber 
\end{align}
\begin{align}
C_{SLR}^{(1)}(m_{\tilde g})&=\frac{\alpha _s^2}{m_{\tilde g}^2}\sum _{I,J=1}^6
\Bigg \{ (\lambda _{GLR}^{(d)})_I^{ij}(\lambda _{GRL}^{(d)})_J^{ij}
\left (-\frac{11}{9}\right )g_{2[1]}(x_I^{\tilde g},x_J^{\tilde g}) \nonumber \\
&\hspace{2cm}+(\lambda _{GLL}^{(d)})_I^{ij}(\lambda _{GRR}^{(d)})_J^{ij}
\left [\frac{14}{3}g_{1[1]}(x_I^{\tilde g},x_J^{\tilde g})
-\frac{2}{3}g_{2[1]}(x_I^{\tilde g},x_J^{\tilde g})\right ]\Bigg \} ,\nonumber \\
C_{SLR}^{(2)}(m_{\tilde g})&=\frac{\alpha _s^2}{m_{\tilde g}^2}\sum _{I,J=1}^6
\Bigg \{ (\lambda _{GLR}^{(d)})_I^{ij}(\lambda _{GRL}^{(d)})_J^{ij}
\left (-\frac{5}{3}\right )g_{2[1]}(x_I^{\tilde g},x_J^{\tilde g}) \nonumber \\
&\hspace{2cm}+(\lambda _{GLL}^{(d)})_I^{ij}(\lambda _{GRR}^{(d)})_J^{ij}
\left [\frac{2}{9}g_{1[1]}(x_I^{\tilde g},x_J^{\tilde g})
+\frac{10}{9}g_{2[1]}(x_I^{\tilde g},x_J^{\tilde g})\right ]\Bigg \} ,
\end{align}
where
\begin{align}
(\lambda _{GLL}^{(d)})_K^{ij}&=(\Gamma _{GL}^{(d)\dagger })_i^K(\Gamma _{GL}^{(d)})_K^j~,\quad 
(\lambda _{GRR}^{(d)})_K^{ij}=(\Gamma _{GR}^{(d)\dagger })_i^K(\Gamma _{GR}^{(d)})_K^j~,\nonumber \\
(\lambda _{GLR}^{(d)})_K^{ij}&=(\Gamma _{GL}^{(d)\dagger })_i^K(\Gamma _{GR}^{(d)})_K^j~,\quad 
(\lambda _{GRL}^{(d)})_K^{ij}=(\Gamma _{GR}^{(d)\dagger })_i^K(\Gamma _{GL}^{(d)})_K^j~.
\end{align}
Here we take $(i,j)=(1,3),~(2,3),~(1,2)$ which correspond to $B^0$, $B_s$, and $K^0$ mesons, respectively. 
The loop functions are given as follows:
\begin{itemize}
\item If $x_I^{\tilde g}\not =x_J^{\tilde g}$ ($x_{I,J}^{\tilde g}=m_{\tilde d_{I,J}}^2/m_{\tilde g}^2$),
\begin{align}
g_{1[1]}(x_I^{\tilde g},x_J^{\tilde g})&=\frac{1}{x_I^{\tilde g}-x_J^{\tilde g}}
\left (\frac{x_I^{\tilde g}\log x_I^{\tilde g}}{(x_I^{\tilde g}-1)^2}
-\frac{1}{x_I^{\tilde g}-1}-\frac{x_J^{\tilde g}\log x_J^{\tilde g}}{(x_J^{\tilde g}-1)^2}
+\frac{1}{x_J^{\tilde g}-1}\right ),\nonumber \\
g_{2[1]}(x_I^{\tilde g},x_J^{\tilde g})&=\frac{1}{x_I^{\tilde g}-x_J^{\tilde g}}
\left (\frac{(x_I^{\tilde g})^2\log x_I^{\tilde g}}{(x_I^{\tilde g}-1)^2}
-\frac{1}{x_I^{\tilde g}-1}-\frac{(x_J^{\tilde g})^2\log x_J^{\tilde g}}{(x_J^{\tilde g}-1)^2}
+\frac{1}{x_J^{\tilde g}-1}\right ).
\end{align}
\item If $x_I^{\tilde g}=x_J^{\tilde g}$,
\begin{align}
g_{1[1]}(x_I^{\tilde g},x_I^{\tilde g})&=
-\frac{(x_I^{\tilde g}+1)\log x_I^{\tilde g}}{(x_I^{\tilde g}-1)^3}+\frac{2}{(x_I^{\tilde g}-1)^2}~,\nonumber \\
g_{2[1]}(x_I^{\tilde g},x_I^{\tilde g})&=
-\frac{2x_I^{\tilde g}\log x_I^{\tilde g}}{(x_I^{\tilde g}-1)^3}+\frac{x_I^{\tilde g}+1}{(x_I^{\tilde g}-1)^2}~.
\end{align}
\end{itemize}
 Taking account of the case that  the gluino mass is  much smaller than the squark mass scale $Q_0$,
the effective Wilson coefficients are given 
by using the RGEs for higher-dimensional operators in Eq.(\ref{Lagrangian-DeltaF=2})
at the leading order of QCD as follows:
\begin{align}
C_{VLL}(m_b(\Lambda =2~\text{GeV}))=&\eta _{VLL}^{B(K)}C_{VLL}(Q_0),\quad 
C_{VRR}(m_b(\Lambda =2~\text{GeV}))=\eta _{VRR}^{B(K)}C_{VLL}(Q_0),\nonumber \\
\begin{pmatrix}
C_{SLL}^{(1)}(m_b(\Lambda =2~\text{GeV})) \\
C_{SLL}^{(2)}(m_b(\Lambda =2~\text{GeV}))
\end{pmatrix}&=
\begin{pmatrix}
C_{SLL}^{(1)}(Q_0) \\
C_{SLL}^{(2)}(Q_0)
\end{pmatrix}X_{LL}^{-1}\eta _{LL}^{B(K)}X_{LL},\nonumber \\
\begin{pmatrix}
C_{SRR}^{(1)}(m_b(\Lambda =2~\text{GeV})) \\
C_{SRR}^{(2)}(m_b(\Lambda =2~\text{GeV}))
\end{pmatrix}&=
\begin{pmatrix}
C_{SRR}^{(1)}(Q_0) \\
C_{SRR}^{(2)}(Q_0)
\end{pmatrix}X_{RR}^{-1}\eta _{RR}^{B(K)}X_{RR},\nonumber \\
\begin{pmatrix}
C_{SLR}^{(1)}(m_b(\Lambda =2~\text{GeV})) \\
C_{SLR}^{(2)}(m_b(\Lambda =2~\text{GeV}))
\end{pmatrix}&=
\begin{pmatrix}
C_{SLR}^{(1)}(Q_0) \\
C_{SLR}^{(2)}(Q_0)
\end{pmatrix}X_{LR}^{-1}\eta _{LR}^{B(K)}X_{LR},
\end{align}
where 
\begin{align}
&\eta _{VLL}^B=\eta _{VRR}^B=\left (\frac{\alpha _s({Q_0})}{\alpha _s(\tilde g)}\right )^{\frac{6}{15}}
\left (\frac{\alpha _s(m_{\tilde g})}{\alpha _s(m_t)}\right )^{\frac{6}{21}}
\left (\frac{\alpha _s(m_t)}{\alpha _s(m_b)}\right )^{\frac{6}{23}},\nonumber \\
&\eta _{LL}^B=\eta _{RR}^B=
S_{LL}
\begin{pmatrix}
\eta _{b\tilde g}^{d_{LL}^1} & 0 \\
0 & \eta _{b\tilde g}^{d_{LL}^2}
\end{pmatrix}
S_{LL}^{-1},\qquad 
\eta _{LR}^B=S_{LR}
\begin{pmatrix}
\eta _{b\tilde g}^{d_{LR}^1} & 0 \\
0 & \eta _{b\tilde g}^{d_{LR}^2}
\end{pmatrix}
S_{LR}^{-1},\nonumber \\
&\eta _{b\tilde g}=\left (\frac{\alpha _s({Q_0})}{\alpha _s(m_{\tilde g})}\right )^{\frac{1}{10}}
\left (\frac{\alpha _s(m_{\tilde g})}{\alpha _s(m_t)}\right )^{\frac{1}{14}}
\left (\frac{\alpha _s(m_t)}{\alpha _s(m_b)}\right )^{\frac{3}{46}},\nonumber \\
&\eta _{VLL}^K=\eta _{VRR}^K=\left (\frac{\alpha _s({Q_0})}{\alpha _s (m_{\tilde g})}\right )^{\frac{6}{15}}
\left (\frac{\alpha _s(m_{\tilde g})}{\alpha _s(m_t)}\right )^{\frac{6}{21}}
\left (\frac{\alpha _s(m_t)}{\alpha _s(m_b)}\right )^{\frac{6}{23}}
\left (\frac{\alpha _s(m_b)}{\alpha _s(\Lambda =2~\text{GeV})}\right )^{\frac{6}{25}},\nonumber
\end{align}
\begin{align}
&\eta _{LL}^K=\eta _{RR}^K=
S_{LL}
\begin{pmatrix}
\eta _{\Lambda \tilde g}^{d_{LL}^1} & 0 \\
0 & \eta _{\Lambda \tilde g}^{d_{LL}^2}
\end{pmatrix}
S_{LL}^{-1},\qquad 
\eta _{LR}^K=
S_{LR}
\begin{pmatrix}
\eta _{\Lambda \tilde g}^{d_{LR}^1} & 0 \\
0 & \eta _{\Lambda \tilde g}^{d_{LR}^2}
\end{pmatrix}
S_{LR}^{-1},\nonumber \\
&\eta _{\Lambda \tilde g}=\left (\frac{\alpha _s({Q_0})}{\alpha _s(m_{\tilde g})}\right )^{\frac{1}{10}}
\left (\frac{\alpha _s(m_{\tilde g})}{\alpha _s(m_t)}\right )^{\frac{1}{14}}
\left (\frac{\alpha _s(m_t)}{\alpha _s(m_b)}\right )^{\frac{3}{46}}
\left (\frac{\alpha _s(m_b)}{\alpha _s(\Lambda =2~\text{GeV})}\right )^{\frac{3}{50}},\nonumber \\
&d_{LL}^1=\frac{2}{3}(1-\sqrt{241}),\qquad d_{LL}^2=\frac{2}{3}(1+\sqrt{241}),\qquad 
d_{LR}^1=-16,\qquad d_{LR}^2=2,\nonumber \\
&S_{LL}=
\begin{pmatrix}
\frac{16+\sqrt{241}}{60} & \frac{16-\sqrt{241}}{60} \\
1 & 1
\end{pmatrix},\quad 
S_{LR}=
\begin{pmatrix}
-2 & 1 \\
3 & 0
\end{pmatrix},\nonumber \\
&X_{LL}=X_{RR}=
\begin{pmatrix}
1 & 0 \\
4 & 8
\end{pmatrix},\qquad 
X_{LR}=
\begin{pmatrix}
0 & -2 \\
1 & 0
\end{pmatrix}.\nonumber \\
\end{align}

%%%%%%%%%%%%%%%%%%%%%%%%%%%%%%%%%%%%%%%%%%%%%%%%%
%%%%% Parameters %%%%%%%%%%%%%%%%%%%%%%%%%%%%%%%%
%%%%%%%%%%%%%%%%%%%%%%%%%%%%%%%%%%%%%%%%%%%%%%%%%

For the parameters $B_i^{(d)}(i=2-5)$ of $B$ mesons,  we use values in  \cite{Becirevic:2001xt}
as follows:
\begin{eqnarray}
&&B_2^{(B_d)} (m_b)=0.79(2)(4), \qquad
B_3^{(B_d)} (m_b)=0.92(2)(4), \nonumber \\
&&B_4^{(B_d)} (m_b)=1.15(3)(^{+5}_{-7}), \qquad 
B_5^{(B_d)} (m_b)=1.72(4)(^{+20}_{-6}), \nonumber\\
&&B_2^{(B_s)} (m_b)=0.80(1)(4), \qquad
B_3^{(B_s)} (m_b)=0.93(3)(8), \nonumber\\
&&B_4^{(B_s)} (m_b)=1.16(2)(^{+5}_{-7}), \qquad
B_5^{(B_s)} (m_b)=1.75(3)(^{+21}_{-6})\ .
\end{eqnarray}
On the other hand, we use the most updated values for $\hat B_1^{(d)} $ and
 $\hat B_1^{(s)} $ as \cite{UTfit}:
\begin{equation}
\hat B_1^{(B_s)}  = 1.33\pm 0.06 \ , \qquad  %KEKFFlattice result  Marco Ciuchini \\
\hat B_1^{(B_s)} / \hat B_1^{(B_d)}=1.05\pm 0.07 \ . % KEKFFlattice result \\
\end {equation}

For the paremeters $B_i^{K}(i=2-5)$, we use following values 
  \cite{Allton:1998sm},
\begin{equation}
\begin{split}
B_2^{(K)}(2{\rm GeV})=0.66\pm 0.04 , \qquad 
B_3^{(K)}(2{\rm GeV})=1.05\pm 0.12, \\
B_4^{(K)}(2{\rm GeV})=1.03\pm 0.06, \qquad
B_5^{(K)}(2{\rm GeV})=0.73\pm 0.10,
\end{split}
\end{equation}
and we take recent value of Eq.(\ref{BK}) for deriving $B_1^{(K)}(2{\rm GeV})$.

For the paremeters $B_i^{D}(i=1-5)$, we use following values  \cite{Buras:2000if,Carrasco:2014uya},
\begin{eqnarray}
&&B_1^{(D)}(3{\rm GeV})=0.75\pm 0.02 , \quad 
B_2^{(D)}(3{\rm GeV})=0.66\pm 0.02 , \quad 
B_3^{(D)}(3{\rm GeV})=0.96\pm 0.05, \nonumber\\
&&B_4^{(D)}(3{\rm GeV})=0.91\pm 0.04, \quad
B_5^{(D)}(3{\rm GeV})=1.10\pm 0.05.
\end{eqnarray}

%%%%%%%%%%%%%%%%%%%%%%%%%%%%%%%%%%%%%%%%%%%%%%%%%%%%%%%%
\section*{Appendix C : The loop functions $F_i$}

The loop functions $F_i(x_{\tilde g}^I)$ are given in terms of  $x_{\tilde g}^I=m_{\tilde g}^2/m_{\tilde d_I}^2~(I=3,6)$
 as follows:
\begin{align}
F_1(x_{\tilde g}^I)&=\frac{x_{\tilde g}^I\log x_{\tilde g}^I}{2(x_{\tilde g}^I-1)^4}
+\frac{(x_{\tilde g}^I)^2-5x_{\tilde g}^I-2}{12(x_{\tilde g}^I-1)^3}~, \quad
F_2(x_{\tilde g}^I)=-\frac{(x_{\tilde g}^I)^2\log x_{\tilde g}^I}{2(x_{\tilde g}^I-1)^4}
+\frac{2(x_{\tilde g}^I)^2+5x_{\tilde g}^I-1}{12(x_{\tilde g}^I-1)^3}~,\nonumber \\
F_3(x_{\tilde g}^I)&=\frac{\log x_{\tilde g}^I}{(x_{\tilde g}^I-1)^3}
+\frac{x_{\tilde g}^I-3}{2(x_{\tilde g}^I-1)^2}~, \quad
F_4(x_{\tilde g}^I)=-\frac{x_{\tilde g}^I\log x_{\tilde g}^I}{(x_{\tilde g}^I-1)^3}+
\frac{x_{\tilde g}^I+1}{2(x_{\tilde g}^I-1)^2}=\frac{1}{2}g_{2[1]}(x_{\tilde g}^I,x_{\tilde g}^I)~.
\end{align}

%%%%%%%%%%%%%%%%%%%%%%%%%%%%%%%%%%%%%%%%%%%%%%%%%%%
\section*{Appendix D : EDM and Chromo-EDM of quarks}
We present 
the EDM of the strange quark from the gluino contribution as the typical example \cite{GotoNote}:
\begin{equation}
d_s(Q_0)=-2\sqrt{4\pi \alpha (m_{\tilde g})}\text{Im}[A_s^{\gamma 22}(Q_0)],
\end{equation}
where 
\begin{align}
A_s^{\gamma 22}&(Q_0)
=\frac{Q_s\alpha _s(m_{\tilde g})}{4\pi }\frac{8}{3}
\sum_{I=1}^6 \frac{1}{2m_{\tilde d_I}^2}\bigg \{ 
\Big (m_s(\lambda _{GLL}^{(d)})_3^{22}+m_s(\lambda _{GRR}^{(d)})_I^{22}\Big )
\Big (F_2(x_{\tilde g}^I)\Big )\nonumber \\
&\hspace{3.3cm}+m_{\tilde g}(\lambda _{GLR}^{(d)})_I^{22}
\Big (F_4(x_{\tilde g}^I)\Big )\bigg \} \ .
\end{align}
On the other hand, 
the chromo-EDM (cEDM) of the strange quark from gluino contribution is given as:
\begin{equation}
d_s^C(Q_0)=-2\sqrt{4\pi \alpha _s(m_{\tilde g})}\text{Im}[A_s^{g22}(Q_0)],
\end{equation}
where 
\begin{align}
%A_s^{g22}&=-\frac{\alpha _s(m_{\tilde g})}{24\pi }
A_s^{g22}&(Q_0)
=-\frac{\alpha _s(m_{\tilde g})}{4\pi }\frac{1}{3}
\sum_{I=1}^6\frac{1}{2m_{\tilde d_I}^2}\bigg \{ 
\Big (m_s(\lambda _{GLL}^{(d)})_I^{22}+m_s(\lambda _{GRR}^{(d)})_I^{22}\Big )
\Big (9F_1(x_{\tilde g}^I)+F_2(x_{\tilde g}^I)\Big )\nonumber \\
&\hspace{3.3cm}+m_{\tilde g}(\lambda _{GLR}^{(d)})_I^{22}
\Big (9F_3(x_{\tilde g}^I)+F_4(x_{\tilde g}^I)\Big )\bigg \} \\
\end{align}

Including the RGE effect of  QCD   \cite{Degrassi:2005zd}, 
the cEDM of the strange quark is given as
 \begin{equation}
d_s^C(2{\rm GeV})=d_s^C(Q_0)
\left ( \frac{\alpha_s(Q_0)}{\alpha_s(m_t)} \right )^{\frac{14}{21}}
 \left ( \frac{\alpha_s(m_t)}{\alpha_s(m_b)} \right )^{\frac{14}{23}}
\left ( \frac{\alpha_s(m_b)}{\alpha_s(2{\rm GeV})} \right )^{\frac{14}{25}} \ .
\end{equation}
On the other hand, the EDM operator is mixied with the cEDM operator during RGE evolution.
Then, one obtains
 \begin{eqnarray}
&&d_s(2{\rm GeV})=d_s(Q_0)
\left ( \frac{\alpha_s(Q_0}{\alpha_s(m_t)} \right )^{\frac{16}{21}}
 \left ( \frac{\alpha_s(m_t)}{\alpha_s(m_b)} \right )^{\frac{16}{23}}
\left ( \frac{\alpha_s(m_b)}{\alpha_s(2{\rm GeV})} \right )^{\frac{16}{25}} 
+\frac{8}{g_s} d_s^C(Q_0) \times \\
&&\left[ \left ( \frac{\alpha_s(Q_0}{\alpha_s(m_t)} \right )^{\frac{16}{21}}
 \left ( \frac{\alpha_s(m_t)}{\alpha_s(m_b)} \right )^{\frac{16}{23}}
\left ( \frac{\alpha_s(m_b)}{\alpha_s(2{\rm GeV})} \right )^{\frac{16}{25}}
-\left ( \frac{\alpha_s(Q_0}{\alpha_s(m_t)} \right )^{\frac{14}{21}}
 \left ( \frac{\alpha_s(m_t)}{\alpha_s(m_b)} \right )^{\frac{14}{23}}
\left ( \frac{\alpha_s(m_b)}{\alpha_s(2{\rm GeV})} \right )^{\frac{14}{25}} \right ].
\nonumber
\end{eqnarray}

The EDMs and  cEDMs of the down- and up-quarks induced by the  gluino interaction are also given by
the similar formulas.

\newpage
%%%%%%%%%%%%%%%%%%%%%%%%%%%%%%%%%%%%%%%%%%%%%%%
%%%%%%%%  Regerences %%%%%%%%%%%%%%%%%%%%%%%%%%
%%%%%%%%%%%%%%%%%%%%%%%%%%%%%%%%%%%%%%%%%%%%%%%

\end{document}